\documentclass[
]{revtex4-1}
\usepackage{graphicx}
\usepackage{dcolumn}
\usepackage{bm}

\usepackage{amssymb}
\usepackage{amsfonts}
\usepackage{amsmath}

\usepackage{graphicx}
\usepackage{subfigure}
\usepackage[colorlinks,linkcolor=red,anchorcolor=blue,citecolor=green]{hyperref}
\usepackage{float}

\begin{document}


\title{Generation of higher-order rogue waves  from multi-breathers\\ by double degeneracy in an optical fibre
}

\author{Lihong Wang$^{1,2}$}
\author{Jingsong He$^{3}$}%
\email{Corresponding author, email: hejingsong@nbu.edu.cn, jshe@ustc.edu.cn}
\author{Hui Xu$^{1}$}%
\author{Ji Wang$^{1}$}%
\affiliation{%
$^{1}$School of Mechanical Engineering $\&$ Mechanics, Ningbo University, Ningbo,  P.\ R.\ China \\
$^{2}$ State Key Laboratory of Satellite Ocean Environment Dynamics \\ (Second Institute of Oceanography, SOA), P.\ R.\ China \\
$^{3}$Department of Mathematics, Ningbo University, Ningbo, Zhejiang 315211, P.\ R.\ China
}%

\author{Kuppuswamy Porsezian }
\affiliation{
 Department of Physics, Pondicherry University, Puducherry 605014, India
}%



\date{\today}

\begin{abstract}

In this paper, we  construct a special kind of breather solution of the nonlinear Schr\"{o}dinger (NLS) equation, the so-called breather-positon ({\it b-positon} for short), which can be obtained by taking the limit $\lambda_{j}$ $\rightarrow$ $\lambda_{1}$ of the Lax pair eigenvalues in the order-$n$ periodic solution which is generated by   the $n$-fold Darboux transformation  from a special ``seed'' solution--plane wave.  Further, an order-$n$ {\it b-positon} gives an order-$n$ rogue wave under a limit $\lambda_1\rightarrow \lambda_0$. Here $\lambda_0$ is a special eigenvalue in a breather of the NLS equation such that its period goes to infinity. Several analytical plots of order-2 breather confirm visually this double degeneration.  The last limit in this double degeneration can be realized approximately in an optical fiber governed by the NLS equation, in which an injected initial ideal pulse  is  created  by a frequency comb system and  a programable optical filter (wave shaper) according to the profile of an analytical form of the {\it b-positon} at a certain position $z_0$.   We also suggest a new way to observe higher-order rogue waves generation in an  optical fiber, namely, measure the patterns at the central region of the higher-order {\it b-positon} generated by above ideal initial pulses  when $\lambda_1$ is very close to the $\lambda_0$. The excellent agreement between the numerical solutions generated from initial ideal inputs with a low signal noise ratio and analytical solutions of order-2 {\it b-positon}, supports strongly this way in a realistic optical fiber system.  Our results also show the validity of the generating mechanism of a higher-order rogue waves from a multi-breathers through the double degeneration.
\begin{description}
\item[PACS numbers]
42.65.Tg, 42.81.Dp, 05.45.Yv, 02.30.Ik
\end{description}
\end{abstract}

\maketitle


\section{Introduction}

During the last fifty years or so, it has been well established that the nonlinear effects are responsible for many exciting inventions. In particular, after the invention of several new mathematical methods and use of supercomputers with the help of advanced software's, we observed an explosive growth and many new concepts in nonlinear science and several new nonlinear evolutions equations have been derived in different branches of science. One of the remarkable solutions admitted by nonlinear partial dispersive equations is the soliton type highly localized solutions. In this work, we consider one such commonly known model of a dispersive nonlinear medium with the cubic self-focusing nonlinearity, which is described, both in optics \cite{chiao}-%
\cite{Hasegawa2} and in general \cite{Zakharov,ablowitzbook}, by the
ubiquitous nonlinear Schr\"{o}dinger (NLS) equation for amplitude $ q $ of the field envelope:
\begin{equation}
\label{gnlse}
\frac{\partial q}{\partial z} +\frac{i\beta_2}{2}\frac{\partial^{2} q}{\partial t^{2}}-i\gamma |q |^2 q =0,
\end{equation}
\noindent where $q(z,t)$ is the envelope amplitude of the electric field at position $z$ in the system, and at time $t$ in the moving frame.  The parameters $\beta_2$ and $\gamma$ designate the chromatic dispersion and Kerr nonlinearity coefficients, respectively. The NLS equation is one of the well-known completely integrable nonlinear systems responsible for many technological developments. Though this equation is well and widely studied in different branches of science, after the introduction and derivation of rogue wave type rational solutions, a lot of re-research has been started on NLS equation, which is evidenced through a large number of publications in the recent past. This equation gives rise to commonly known solitons, which were experimentally generated in nonlinear optical fibers as temporal pulses \cite{mollenauer}, and in planar waveguide as self-trapped beams \cite{spatial-sol}. The NLS equation has also been derived in many branches of physics. From the detailed investigations of this equation during the past forty years or so, it is well known that NLS equation admits many types of solutions like solitons, breathers, similaritons, etc. A variety of optical solitons were studied in detail through theoretical and experimental techniques in NLS type equations, including spatiotemporal solitons confined in both space and time, solitary vortices, the Bragg solitons mentioned above and those supported by non-Kerr nonlinearities (such as quadratic), discrete and lattice self-trapped modes, breathers, dissipative solitons, etc. (see, in particular, reviews \cite{R1}-\cite{R10} and books \cite{Hasegawa3}-\cite{agrawalbook5ed}). Further, this equation is responsible for many recent technological developments in the area of modern nonlinear optical fibre namely many soliton trials in commercial networks, pedestal free pulse compression, soliton laser, soliton based supercontinuum generation, etc. Though this equation needs modification depends on the nature and power of the pulse transmission through optical fibre, in this paper, we restrict our discussion to standard NLS equation without any additional perturbation. This is mainly because, in this work, our prime aim is to report the new type of solutions in the form of breather-positons ({\it b-positons} for short and will be defined later) and the generation of higher-order rogue waves from multiple breathers.

 In recent years, a doubly localized solution both in space and in time dimensions, i.e. rogue wave, of the NLS equation has been extremely attracted much attentions from  theoretical and experimental concerns,  although this kind of qusi-rational solution  \cite{Peregrine,Akhmedievrw1985}  was reported more than 30 years ago by a simple limit of a traveling periodic solution (breather) \cite{Kuznetsovbreather,Mabreather,Abreather}.  By comparing with rich observations  for  the different-order rogue waves  of the NLS equation in water tank \cite{amin2011watertank, amin2012watertankprx,amin2012watertankpre,amin2013watertankpla, amin2014watertankprl}, the laboratory works in optical system are  subjected only  to the order-1 rogue wave \cite{Kibler1,Kibler2,Kibler3,Kibler4}.  The order-1 rogue wave is a limit of a breather when its period goes to infinity  \cite{Peregrine,Akhmedievrw1985},  thus it can be  represented approximately by a peak of the breather when the period of the breather is sufficiently large. Indeed,   the first observation of the order-1 rogue wave has been realized in a nonlinear fiber by  creation of a breather after injecting a modulation wave \cite{Kibler1}, in which  a key parameter $a$ is adjustable through fine tuning  of  the initial power $P_0$ and frequency $\omega_{mod}$.  Soon after, this observation has been implemented again in standard SMF-28 fiber \cite{Kibler2}. Note that a breather will convert into a order-1 rogue wave when parameter $a\rightarrow 0.5$.  In the above mentioned optical experiments,  $a=0.42$  \cite{Kibler1} and $a=0.47$  \cite{Kibler2}, and thus one peak of a breather in two experiments is an excellent approximation of the order-1 rogue wave  of the NLS.

However, it is a quite challenging task to observe higher-order rogue waves in a fiber. To date, several methods have been proposed to generate higher-order rogue waves\cite{hehrwpre2013,akhmedievcirularrwpre2011}. In principle, the higher-order rogue waves are generated from a multi-breathers by double degeneration\cite{hehrwpre2013,akhmedievcirularrwpre2011}. The first degeneration of an order-$n$  breather is the limit of eigenvalues $\lambda_i\rightarrow \lambda_1$,  and the second degeneration is the limit $\lambda_1\rightarrow \lambda_0$.   Here $\lambda_0$ is a special eigenvalue such that the period of a breather of the NLS goes to infinity, and then this breather converts into an order-1 rogue wave.
Note that the double degeneration is also expressed by a similar way as two steps through modulation frequency: first  $\kappa_j=j\kappa$ and then $\kappa \rightarrow 0$ \cite{akhmedievcirularrwpre2011}, and the frequency is expressed by an imaginary eigenvalue $l_j$ as $\kappa_j=2\sqrt{1+l_j^2}$.  In other words, higher-order rogue waves can be generated from the collision of several breathers, and this has been numerically and approximately demonstrated through profiles of the intersection area of them \cite{hehrwpre2013,akhmedievpla2009nonzerovecolitybreather,akhmedievpra2009excite,akhmedievpra2013preclassifying}. Using  the progressive fussion and fission of $n$ degenerate breathers associated with a critical eigenvalue creates an order-n rogue wave. Through this mechanism, we also proved two important conjectures regarding the total number of peaks and decomposition rule in the circular pattern of an order-n rogue wave \cite{hehrwpre2013}. For example, figures \ref{fig1}-\ref{fig3} of ref. \cite{hehrwpre2013} provide approximately three patterns of the order-3 rogue waves by using three different eigenvalues.  However, as shown in ref. \cite{akhmedievpla2009nonzerovecolitybreather,akhmedievpra2009excite}, it is very difficult to control the ``velocity"(or equivalently called ``period") and phase of the different  breathers to realize the effective collision, and then get approximately the different patterns of the higher-order rogue waves in the strong intersection area. More specifically, one cannot use initial power $P_0$ and frequency $\omega_{mod}$  to realize  approximately  the transferring between multi-breathers and  higher-order rogue waves, because there are different periods (or equivalent frequencies) for different breathers, unlike a single breather just has one period  which can be adjusted effectively in experiments \cite{Kibler1,Kibler2,Kibler3,Kibler4}.  The initial field to create multi-breathers is a superposition of several exponential functions on the top of a plane wave \cite{akhmedievpra2009excite}, which is a main source of the above difficult point to produce effective collision of breathers. Thus from these studies, it is quite clear that, because of tedious mathematical steps and several possible patterns, it is quite tricky and challenging to generate the higher-order rogue waves.

 The purposes of this paper are, i) to show theoretically the two steps of the double degeneration from multi-breathers to higher-order rogue waves of the NLS equation, and ii) to suggest a new way to observe above-mentioned higher-order rogue waves in a fiber.   To this end, we introduce a special kind of breather, the {\it b-positon}, which is obtained by taking the limit $\lambda_{j}$ $\rightarrow$ $\lambda_{1}$ of the Lax pair eigenvalues in the order-$n$ periodic solution which is generated by  the $n$-fold Darboux transformation (DT)  from a plane wave. Further, an order-$n$ {\it b-positon} gives an order-$n$ rogue wave under a limit $\lambda_1\rightarrow \lambda_0$.  This type of generating mechanism of rouge waves has been explained clearly in ref. \cite{hehrwpre2013}.  The order-$n$ {\it b-positon} is a transmission state between order-$n$ breather to an order-$n$ rogue wave, in which different breathers have same period ( or velocity) and can have (or have not) different phases, and they  produces different patterns in the strong interaction area.  In fact, an order-1 {\it b-positon} is a single breather. Moreover, an explicit form of a order-2 Akhmediev breather with two different modulation frequencies (or equivalently two imaginary eigenvalues) and shifts has been given in ref. \cite{akhmedievpre2013pre2ndabs}, and the degenerate Akhmediev breather (see eq. (7) of this reference) by a limit of equal eigenvalues has also been provided explicitly, which is a special order-2 {\it b-positon}. There are two distinct points in our paper, by comparing with ref. \cite{akhmedievpre2013pre2ndabs}: 1) eigenvalues are not imaginary, which results in propagation of the {\it b-positon} in an arbitrary direction besides two axes; 2) We use parameters $s_i$ (see details in the following section) instead of shifts to control the phases of the breathers, which  is more convenient to generate more complex and interesting patterns.

The rest of the paper is organized as follows. After detailed introduction, method of derivation of breather solutions is given in section II. In section III, we report the derivation of the {\it b-positon} solutions.  The observation of higher-order rogue waves and {\it b-positons} in an optical fibre is discussed in section IV. Finally, the numerical simulations of the {\it b-positon} are demonstrated in section V  and results are summarized in section VI.

\section{Higher-order breather solutions}

To obtain explicit forms of the higher-order breathers, we set  $\beta_{2}=-2$ and $\gamma =2$ in
 Eq. (\ref{gnlse}) for our further discussion.   We shall use determinant representation \cite{matveev, hedarbroux}  of the $n$-fold Darboux  transformation to get breathers of the NLS equation.  Based on our early results\cite{hehrwpre2013,hebreather1,hebreather2}, we start with a special kind of  ``seed'' solution -- plane wave,
\begin{equation}\label{q0}
q^{[0]} =c e^{i \rho},
\end{equation}
in which $\rho= a t + \left( 2c^2-a^2 \right) z,  a,c \in \mathbb{R}, c\neq 0$. The eigenfunction \cite{hehrwpre2013,hebreather1,hebreather2} associated with eigenvalue $\lambda$ and above seed solution  is
\begin{equation}\label{phi}
\phi(\lambda)=
\left(
\begin{array}{ll}
c e^{ i \left( \frac{\rho}{2} +  d(\lambda)\right)}   +i \left(\frac{a}{2} + \lambda + h(\lambda) \right) e^{ i \left(\frac{\rho}{2} - d(\lambda) \right)}   \\
c e^{-i \left(\frac{\rho}{2}  +  d(\lambda)\right)}  +i \left(\frac{a}{2} + \lambda + h(\lambda) \right) e^{-i \left(\frac{\rho}{2} - d(\lambda) \right)}
\end{array}
\right).
\end{equation}
Here $ h(\lambda)=\sqrt{c^2+ \left( \lambda+\frac{a}{2} \right)^2} ,  $
$ d(\lambda) =(t+(2\lambda-a)z+S_0)   h(\lambda) ,$
$ S_0 =  s_0+\sum_{k=1}^{n-1} s_k \epsilon^{2k} . $

To get order-$n$ breather \cite{hebreather2} by using determinant representation \cite{hedarbroux}  of  $n$-fold DT, we select eigenvalue and its eigenfunction as follows:
\begin{equation}\label{sec1}
f_{2k-1}=
\left(
\begin{array}{ll}
f_{2k-1,1}(\lambda_{2k-1})\\
f_{2k-1,2}(\lambda_{2k-1})
\end{array}
\right) = \phi(\lambda_{2k-1}) \;  \text{for }  \lambda_{2k-1},
\end{equation}
but
\begin{equation}\label{sec2}
 f_{2k}=
\left(
\begin{array}{ll}
f_{2k,1}(\lambda_{2k})\\
f_{2k,2}(\lambda_{2k})
\end{array}
\right)=
\left(
\begin{array}{rr}
-f_{2k-1,2}^*(\lambda_{2k-1})\\
f_{2k-1,1}^*(\lambda_{2k-1})
\end{array}
\right)\;  \text{for  }  \lambda_{2k}= \lambda_{2k-1}^*.
\end{equation}
Then an order-$n$ breather\cite{hebreather2} of the NLS is formulated as
\begin{equation}\label{qn}
q^{[n]}=q^{[0]}-2i\frac{|\Delta_1^{[n]}|}{|\Delta_2^{[n]}|}.
\end{equation}
Here two matrices are
\begin{widetext}
$$\Delta_1^{[n]}\mbox{\hspace{-0.1cm}}=\mbox{\hspace{-0.2cm}}
\left(\mbox{\hspace{-0.1cm}}\begin{array}{ccccccccc} f_{11}&f_{12}& \lambda_1
f_{11}&\lambda_1 f_{12}&\lambda^2_1f_{11}&\lambda^2_1 f_{12}
 &\mbox{\hspace{-0.4cm}}\cdots\mbox{\hspace{-0.4cm}}&\lambda^{n-1}_1 f_{11} & \lambda^n_1 f_{11}\\
f_{21}&f_{22}& \lambda_2 f_{21}&\lambda_2 f_{22}&\lambda^2_2
f_{21}&\lambda^2_2 f_{22}
 &\mbox{\hspace{-0.4cm}}\cdots\mbox{\hspace{-0.4cm}}&\lambda^{n-1}_2 f_{21}& \lambda^n_2 f_{21}\\
f_{31}&f_{32}& \lambda_3 f_{31}&\lambda_3 f_{32}&\lambda^2_3
f_{31}&\lambda^2_3 f_{32}
 &\mbox{\hspace{-0.4cm}}\cdots\mbox{\hspace{-0.4cm}}&\lambda^{n-1}_3 f_{31}& \lambda^n_3 f_{31}\\
\vdots &\vdots &\vdots &\vdots &\vdots &\vdots &\mbox{\hspace{-0.4cm}}\cdots \mbox{\hspace{-0.4cm}}&\vdots &\vdots\\
f_{2n 1}&f_{2n 2}& \lambda_{2n} f_{2n 1}&\lambda_{2n} f_{2n
2}&\lambda^2_{2n} f_{2n 1}&\lambda^2_{2n} f_{2n 2}
 &\mbox{\hspace{0cm}}\cdots\mbox{\hspace{0cm}}&\lambda^{n-1}_{2n} f_{2n 1}& \lambda^n_{2n} f_{2n 1}
\end{array}\mbox{\hspace{-0.1cm}}\right),
$$

$$\Delta_2^{[n]}\mbox{\hspace{-0.1cm}}=\mbox{\hspace{-0.2cm}}
\left( \mbox{\hspace{-0.1cm}}\begin{array}{ccccccccc}
 f_{11}&f_{12}& \lambda_1 f_{11}&\lambda_1f_{12}&\lambda^2_1 f_{11}&\lambda^2_1 f_{12}
 &\mbox{\hspace{-0.4cm}}\cdots\mbox{\hspace{-0.4cm}}&\lambda^{n-1}_1 f_{11}&\lambda^{n-1}_1 f_{12}\\
f_{21}&f_{22}& \lambda_2 f_{21}&\lambda_2 f_{22}&\lambda^2_2
f_{21}&\lambda^2_2 f_{22}
 &\mbox{\hspace{-0.4cm}}\cdots\mbox{\hspace{-0.4cm}}&\lambda^{n-1}_2 f_{21}&\lambda^{n-1}_2 f_{22}\\
f_{31}&f_{32}& \lambda_3 f_{31}&\lambda_3f_{32}&\lambda^2_3
f_{31}&\lambda^2_3 f_{32}
 &\mbox{\hspace{-0.4cm}}\cdots\mbox{\hspace{-0.4cm}}&\lambda^{n-1}_3 f_{31}&\lambda^{n-1}_3 f_{32}\\
\vdots &\vdots &\vdots &\vdots &\vdots &\vdots &\mbox{\hspace{-0.4cm}}\cdots \mbox{\hspace{-0.4cm}}&\vdots &\vdots\\
f_{2n 1}&f_{2n 2}& \lambda_{2n} f_{2n 1}&\lambda_{2n}f_{2n
2}&\lambda^2_{2n} f_{2n 1}&\lambda^2_{2n} f_{2n 2}
 &\mbox{\hspace{0cm}}\cdots\mbox{\hspace{0cm}}&\lambda^{n-1}_{2n} f_{2n 1}&\lambda^{n-1}_{2n} f_{2n 2}\\
\end{array}
 \mbox{\hspace{-0.1cm}}\right).$$
\end{widetext}
There are two real variables $z,t$, three real parameters $a, c, \epsilon$  and $2n$ complex parameters $\lambda_i(i=1,3,5,\cdots 2n-1),s_i(i=0,1,2,\cdots n-1)$ in
$q^{[n]}$.  In general,   $\epsilon$  is an arbitrary real parameter in $q^{[n]}$, and we usually  set that $S_0$ is a monomial of $\epsilon$  with  $\epsilon=1$, i.e.
$S_0=s_k\epsilon^{2k}|_{\epsilon=1}=s_k$ which controls the central pattern of this breather.
However,  $\epsilon$ is an infinitesimal parameter in $q^{[n]}$ when one  constructs the higher-order {\it b-positons} and rogue waves by the degeneracy  limit of eigenvalues, and its coefficients $s_k$  \cite{hehrwpre2013} are very crucial  to control the pattern of the obtained solutions.

To illustrate this method of the construction of the breather, we would like to provide specific examples.
Set $n=1$ in eq.(\ref{qn}), which gives the one-fold DT, and a new solution
\begin{equation}\label{q1}
q^{[1]}=q^{[0]}- 2i \frac { |\Delta_1^{[1]}| }{ |\Delta_2^{[1]}| },
\end{equation}
$$\text{with\; \; }
\Delta_1^{[1]}=
\left|
\begin {array}{cc}
f_{{1,1}}& \lambda_{{1}}f_{{1,1}
}
\\
f_{{2,1}}& \lambda_{{2}}f_{{2,1}}
\end {array}
\right|=
(\lambda_1^*-\lambda_1) f_{{1,1}}f_{{2,1}},
$$
$$\text{and\; \; }
\Delta_2^{[1]}=
\left|
\begin {array}{cc}
f_{{1,1}}&f_{{1,2}}
\\
f_{{2,1}}&f_{{2,2}}
\end {array}
\right|
=
f_{{1,1}}f_{{2,2}}-f_{{1,2}}f_{{2,1}}.
$$
Substituting  $\lambda_{1}=\xi+i\eta,\lambda_{2}=\lambda_{{1}}^*=\xi-i\eta, S_0=0$, and
$$
f_1=
\left(
\begin{array}{ll}
f_{1,1}(\lambda_1)\\
f_{1,2}(\lambda_1)
\end{array}
\right)=
\phi(\lambda_1),
f_2=
\left(
\begin{array}{ll}
f_{2,1}(\lambda_2)\\
f_{2,2}(\lambda_2)
\end{array}
\right)=
\left(
\begin{array}{rr}
-f_{1,2}^*(\lambda_2^*)\\
f_{1,1}^*(\lambda_2^*)
\end{array}
\right)=
\left(
\begin{array}{rr}
-f_{1,2}^*(\lambda_1)\\
f_{1,1}^*(\lambda_1)
\end{array}
\right),
$$
into eq.(\ref{q1}), after a tedious simplification, we get an explicit formula of the  order-1 breather
{\footnotesize
\begin{widetext}
\begin{equation}\label{1breather}
q^{[1]}=\mbox{\hspace{-0.2cm}}\left(c+\frac
{2\eta\left\{\left[w_1\cos\left(2G\right)-w_2\cosh\left(2F\right)
\right]-i\left[\left(w_1-2c^2\right)\sin\left(2G\right)- w_3\sinh
\left(2F\right)\right]\right\}}{w_1\cosh\left(2F\right)-w_2\cos
\left(2G\right)}\right){\rm e}^{i\rho},
\end{equation}
\end{widetext}
}
\noindent with $w_1={c}^{2}+(h_I+\eta)^{2}+(
\xi+h_R+\frac{a}{2})^{2},w_2=2c(h_I+\eta),
w_3=2c(\xi+h_R+\frac{a}{2})$, $F=  t h_I + d_I z,G=
t h_R + d_R z, d_R=\Re(d), d_I=\Im(d), h_R=\Re(h), h_I=\Im(h)$.  This is a periodic traveling wave.
It is well-known that the order-1 rogue wave is obtained from $q^{[1]}$  by a limit $\lambda_1\rightarrow \lambda_0$, which is given in appendix. Here
$\lambda_0=-\frac{a}{2}+ic$.

Taking $n=2$ in eq.(\ref{qn}), and according to the selections in eqs. (\ref{sec1},\ref{sec2}) of $f_i(i=1,2,3,4),\lambda_2=\lambda_1^*$ and $\lambda_4=\lambda_3^*$,
then an order-2 breather can be expressed by
\begin{eqnarray}\label{2breather}
q^{[2]}=&q^{[2]}(z,t;a,c;\lambda_1,\lambda_3;s_0,s_1,\epsilon)=q^{[0]}-2i\frac{|\Delta_1^{[2]}|}{|\Delta_2^{[2]}|}.
\end{eqnarray}
Here
$$
\Delta_{1}^{[2]}=
\left(
\begin {array}{cccc}
f_{1,1}&f_{1,2}& \lambda_1 f_{1,1} & \lambda_1^2 f_{1,1} \\
f_{2,1}&f_{2,2}& \lambda_2 f_{2,1} & \lambda_2^2 f_{2,1} \\
f_{3,1}&f_{3,2}& \lambda_3 f_{3,1} & \lambda_3^2 f_{3,1} \\
f_{4,1}&f_{4,2}& \lambda_4 f_{4,1} & \lambda_4^2 f_{4,1}
\end {array}
\right),
$$
and
$$
\Delta_{2}^{[2]}=
\left(
\begin {array}{cccc}
f_{1,1}&f_{1,2}& \lambda_1 f_{1,1}& \lambda_1  f_{1,2} \\
f_{2,1}&f_{2,2}& \lambda_2 f_{2,1}& \lambda_2  f_{2,2} \\
f_{3,1}&f_{3,2}& \lambda_3 f_{3,1}& \lambda_3  f_{3,2} \\
f_{4,1}&f_{4,2}& \lambda_4 f_{4,1}& \lambda_4  f_{4,2}
\end {array}
\right).
$$
Using this determinant representation, an explicit form of the order-2 breather has been given in
ref. \cite{hebreather1}.  Recently, a new form of this breather is given in \cite{akhmedievpre2013pre2ndabs}. Moreover, we have used this method to get an order-3 breather, and plotted the central profiles\cite{hehrwpre2013} which provide a good approximation of an order-3 rogue wave with three distinct (minor difference) eigenvalues.

In general,  an order-2 breather $q^{[2]}$ is an indeterminate form  $\frac{0}{0}$ when  $\lambda_3\rightarrow\lambda_1$.  This observation inspires us to consider the limit of $q^{[n]}$ when $\lambda_i \rightarrow \lambda_1$ \cite{hehrwpre2013} from an arbitrary ``seed'' solution $q^{[0]}$.  In the next section, we shall study this degenerate limit of eigenvalues of order-$n$ breather $q^{[n]}$ in eq. (\ref{qn}).

\section{The order-$n$ {\it b-positon} solution}

As we have shown in ref. \cite{hehrwpre2013}, an order-$n$ breather $q^{[n]}$ reduces to an order-$n$ rogue wave  by double degeneration: $\lambda_i\rightarrow \lambda_1$ and then $\lambda_1\rightarrow \lambda_0=-\frac{a}{2}+ic$. However the calculation of this double degeneration is implemented by one step as $\lambda_i=\lambda_0+\epsilon$ with the help of symbolic computational software, and thus ignore the attention of the first limit of $q^{[n]}$.
Here we use this limit to define breather-positon ({\it b-positon} for short): an order-$n$ {\it b-positon}  is  obtained by taking the limit $\lambda_{i}$ $\rightarrow$ $\lambda_{1}$ of the Lax pair eigenvalues in an order-$n$  breather, namely  $q^{[n]}_{\rm{{\it b\mbox{-}positon}}}= \lim_{\lambda_i\rightarrow \lambda_1}q^{[n]}$($\lambda_1\not=\lambda_0$).  Note that  $\lambda_i\rightarrow \lambda_1$ means
 $\lambda_{2i+1}\rightarrow \lambda_1$ and  $\lambda_{2i}\rightarrow \lambda_1^{*}(i=1,2,\cdots,n)$ simultaneously because of the selection $\lambda_{2k}=\lambda_{2k-1}^{*}$ in eq.(\ref{sec2}).   This solution  is  an  extension of the  positon solutions reported  in  Matveev's papers \cite{matveevpostionpla1992-1,matveevpostionpla1992-2} because it denotes  the degeneracy  of multi-solitons under same eigenvalue for KdV and mKdV equations.

     According to the  above definitions, under the limit $\lambda_i \rightarrow \lambda_1$, an indeterminate form $\frac{0}{0}$ associated with $q^{[n]}$ yields an order-$n$ {\it b-positon} by higher-order Taylor expansion, namely
\begin{equation}\label{degenerateddtqn}
    q^{[n]}_{\rm{{\it b\mbox{-}positon}}}(z,t)=q^{[0]}-2i\frac{| \Delta_1^{'[n]}|}{|\Delta_2^{'[n]}|},
\end{equation}
with
\begin{align}
\Delta_1^{'[n]}&=\left(\frac{\partial^{n_i-1}}{\partial\varepsilon^{n_i-1}}\bigg|_{\varepsilon=0}
    ( \Delta_1^{[n]})_{ij} (\lambda_1+\epsilon)\right)_{2n\times2n}, \nonumber \\
\Delta_2^{'[n]}&=\left(\frac{\partial^{n_i-1}}{\partial\varepsilon^{n_i-1}}\bigg|_{\varepsilon=0}
    ( \Delta_2^{[n]})_{ij}
    (\lambda_1+\epsilon)\right)_{2n\times2n},\nonumber
\end{align}
$n_i=[\frac{i+1}{2}]$, $[i]$ denotes the floor function of $i$. It should be noted that  there are two real parameters
$a,c$, and $n+1$ complex parameters $\lambda_1, s_i(i=0,1,2,\cdots,n-1)$ in an order-$n$ {\it b-positon}.   It is trivial to find an order-1 {\it b-positon} which  is a single breather. Furthermore, an order-$n$  {\it b-positon} yields an  order-$n$  rogue wave when $\lambda_1\rightarrow \lambda_0$ \cite{hehrwpre2013}.

The first  nontrivial {\it b-positon} is an order-2 {\it b-positon}. To get a relatively simple form, set $n=2, \lambda_1=\xi_1+i\eta_1$, and $\xi_1=-\frac{a}{2}$ in  eq.(\ref{degenerateddtqn}), then an explicit form of $ q^{[2]}_{\rm{{\it b\mbox{-}positon}}}$ is given in appendix, which is used to plot figure \ref{fig2} with $a=0$. This is similar to the result of ref. \cite{akhmedievpre2013pre2ndabs}. The order-2 rogue wave is also given in appendix through the limit of
 $q^{[2]}_{\rm{{\it b\mbox{-}positon}}}$ as $\lambda_1\rightarrow \lambda_0$ with a condition on  imaginary part of $s_0$, i.e.  $\Im{(s_0)}=0$. However, it is interesting to note that for order-2 and order-3 {\it b-positon}, set  $\xi_1 \not=-\frac{a}{2}$ in  eq.(\ref{degenerateddtqn}), we get tilted propagation of  {\it b-positons}, and their density profiles are presented in appendix (see figure \ref{fig10} and figure \ref{fig30}). As a byproduct, in figure \ref{fig20}, a nonzero real part of the $s_0$ in an order-2 {\it b-positon}  results in  a remarkable shift of the central profile along $t$-axis, but  a nonzero imaginary  part of the $s_0$ produces a shift along $z$-axis with a deformation of central peaks. In addition, the distance between two peaks in figure \ref{fig20}  is increasing significantly along the time axis by comparing with the two pictures in last row of figure \ref{fig10}.

  To  illustrate the further application of the construction for  the order-$n$ {\it b-positon}, set $n=3, \lambda_1=\xi_1+i\eta_1$, and $\xi_1=-\frac{a}{2}$ in  eq.(\ref{degenerateddtqn}), then an explicit form of $ q^{[3]}_{\rm{{\it b\mbox{-}positon}}}$ is given in appendix. This analytical formula of an order-3 {\it b-positon}  is used  to plot figure \ref{fig3}(a)  and figure \ref{fig3xx}(a,b) with parameter   $a=0$. Moreover, due to the appearance of the  $s_i$, $q^{[n]}_{\rm{{\it b\mbox{-}positon}}}$ can generate systematically  different patterns in the central region of the {\it b-positon}, see examples up to order-5 in figures (\ref{fig2}-\ref{fig5}). These patterns resemble very much like corresponding rogue waves  when $\lambda_1$ is close to  $\lambda_0$. Note that, in particular, an order-3 {\it b-positon} which is propagating along  $z$-axis is given by setting ${\rm Im}(\lambda_1) > c$, which is plotted in figure \ref{fig3}(d).

\begin{figure}[H]
  \centering
  \subfigure[]{\includegraphics[scale=.4]{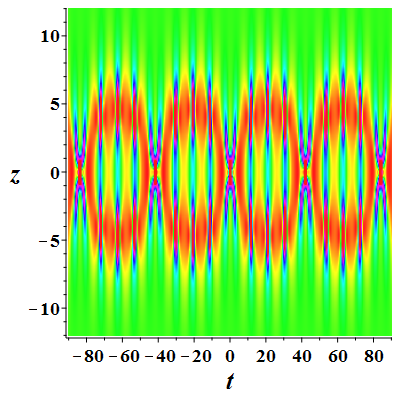}   \label{fig1:a}}
  \subfigure[]{\includegraphics[scale=.4]{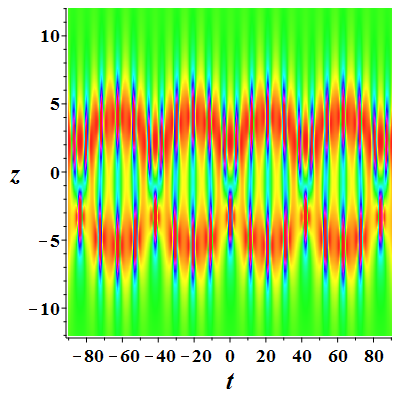}   \label{fig1:b}} \\
  \subfigure[]{\includegraphics[scale=.4]{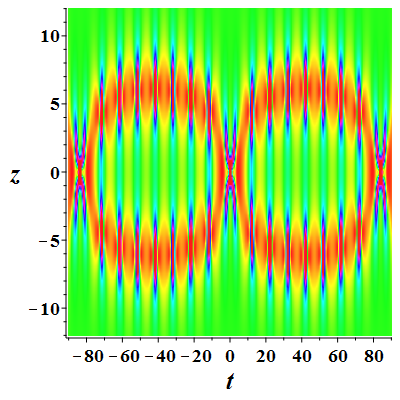}  \label{fig1:c}}
  \subfigure[]{\includegraphics[scale=.4]{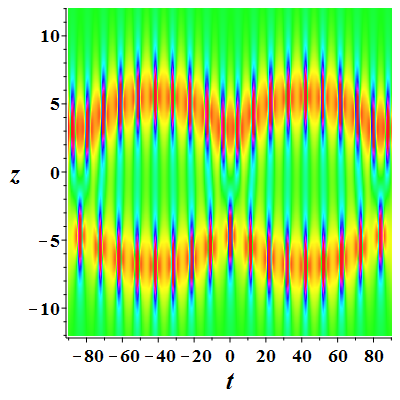}  \label{fig1:d}} \\
  \subfigure[]{\includegraphics[scale=.4]{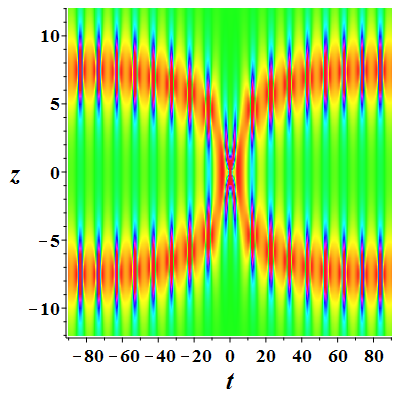}  \label{fig1:e}}
  \subfigure[]{\includegraphics[scale=.4]{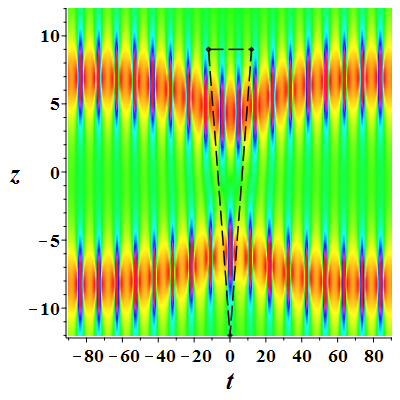}  \label{fig1:f}}
  \caption{(Color Online) A sketchy demonstration of the  limit  $\lambda_3 \rightarrow \lambda_1$ in an   order-2 breather
  $|q^{[2]}|^2$ (density plot) with $a=0,c=\frac{1}{2},\lambda_1=\frac{2}{5}i, s_1=0$. The central region of the left (right) column  is a fundamental (triangular) pattern.   From top to bottom, the ratio of two breathers are 4:5, 8:9, 15:16, and there are 3, 7, 14 peaks in one period of time.
  The other parameters of the breathers respectively are
  (a) $\lambda_3=\frac{\sqrt{7}}{8} i ,s_0=0$,
  (b) $\lambda_3=\frac{\sqrt{7}}{8} i,s_0=1$,
  (c) $\lambda_3=\frac{\sqrt{871}}{80} i,s_0=0$,
  (d) $\lambda_3=\frac{\sqrt{871}}{80} i,s_0=1$,
  (e) $\lambda_3=\frac{3\sqrt{41}}{50} i ,s_0=0$,
  (f) $\lambda_3=\frac{3\sqrt{41}}{50} i ,s_0=1$.
  }
  \label{fig1}
\end{figure}

 We are now in a position to demonstrate intuitively  two limits of double  degeneration  by a graphical way based on analytical  solutions $q^{[2]}$ and  $q^{[2]}_{\rm{{\it b\mbox{-}positon}}}$, i.e. an order-2 breather to an order-2
 {\it b-positon} by $\lambda_3\rightarrow \lambda_1$, and then to an order-2 rogue wave by $\lambda_1\rightarrow \lambda_0$.
\begin{itemize}
\item Figure \ref{fig1} shows that the overall circular  periodic structure gradually enlarges when $\lambda_3$ goes to $\lambda_1$ until  it disappears completely and only one intersection area is preserved, which is quasi-periodic with respect to the peaks in two rows,  when  an order-2 breather becomes an order-2 {\it b-positon}.
\item Figure \ref{fig2} shows that the peaks around the central region gradually leave when $\lambda_1$ goes to $\lambda_0$ until the central profile is survived only and  all other peaks disappear, when an order-2 {\it b-positon}  becomes an order-2 rogue wave.
\end{itemize}
 The difference between the two columns of figures \ref{fig1} and \ref{fig2} is the fundamental pattern(i.e.,one main peak surrounded by several gradually decreasing small peaks on both the sides) or triplet pattern in central region.
 It is a natural request in  two limits of double  degeneration  to use  a same set of parameters $\{s_0,s_1\}$ in the last row  of figure \ref{fig1} and the   first row of figure \ref{fig2}, in order to emphasize the limit process of $\lambda_3\rightarrow \lambda_1\rightarrow \lambda_0$. However  we use different values of them in the right column  in order to get a higher visibility of pictures.

  Four animations are provided in the Supplemental Material \cite{Supplemental_Material} for the analytical demonstration of double degeneracy.
  The animations demonstrate clearly the tendency  from  a  breather to a  {\it b-positon} and  from  a  {\it b-positon} to a rogue wave, which are corresponding to the figure \ref{fig1} and figure \ref{fig2} respectively.

  It is very clear from figure \ref{fig2} that the conversion between {\it b-positon} and the rogue wave is very similar to the transmission between the single breather and the order-1 rogue wave, and the later transmission has been used to observe an  order-1 rogue wave in optical fiber \cite{Kibler1,Kibler2,Kibler3,Kibler4}. Further, in a higher-order {\it b-positon}(see figures \ref{fig2}-\ref{fig5}), the effective collision of multi-breather can be reached,  and the different patterns in central region which are good approximations of the corresponding rogue waves, can be controlled by the $s_i$.  For example, the central profiles in the figures \ref{fig2}(a,c,e) look like fundamental order-2 rogue waves very much, which  is verified by  the excellent coincidence of two pulses in  each panel of figure \ref{twolineatzero}.  Moreover figure \ref{twolineatzero}  also shows that the remarkable decrease in error when $\lambda_1$ is approaching $\lambda_0$. Thus, we can use above two advantages of the {\it b-positon} to observe higher-order rogue wave in an optical fiber.

\section{Observation of  higher-order rogue waves in an optical fiber\\ through the {\it b-positon}}

Although it is difficult to implement effective collision \cite{akhmedievpla2009nonzerovecolitybreather,akhmedievpra2009excite} of two or three  breathers in optical experiment, Frisquet etal. \cite{kilberprx2013collision}  have realized firstly  the collision of two breathers by injecting a  bimodulated continuous wave with two  distinct frequencies, which is expressed  by two exponential functions with two small real amplitudes \cite{kilberprx2013collision}.  However, due to the non-ideal initial pulse in their experiment, there exists a non-ignorable discrepancy of the main peak of synchronized collision in theory and observation, and its difference is almost 4 (for details see figure 7(a) of ref. \cite{kilberprx2013collision}). Of course, the most accurate way to observe collision is to inject an  ideal (and  exact) initial pulse in terms of a certain initial function $q(z_0,t)$ at a suitable position $z_0$  taken from an  exact and analytical solution of breathers $q(z,t)$ for the NLS equation, and then measure the intensity or optical spectrum  of  output pulse at a certain position $z_1$. Unfortunately, this kind of ideal (and  exact) initial pulse  usually has a more  complex profile, which is impossible to be produced  by a common optical signal generator.

Recently,  a frequency comb  and a programable optical filter (wave shaper) (see figure 1 in ref.\cite{kilberpra2014generator}) are used to create an  ideal initial pulse according to the profile of an analytical solution, and  then an order-2 breather has been observed successfully with a  typical X-shape signature in the plane of $(z,t)$ (see figure 3 of ref.\cite{kilberpra2014generator}).  Very recently,  these  powerful devices and technique have  been  used again to observe one  pair of breathers  (see figure 5  in ref.\cite{kilberprx2015superregular}) with same heights but opposite propagating directions (or called a supperregular breather \cite{{zakharovnon2014supper}} ), from an ideal initial pulse. Of course, these results are far away from the order-2 rogue waves of the NLS equation, which are not their actual objectives \cite{kilberprx2015superregular}.

 Based on the two advantages of the {\it b-positon}, i.e., a convenient conversion to the rogue wave and the easy  controllability of the patterns in the central region in the ($z,t$)-plane, which have been pointed out at the  end of section III, we introduce following new way to observe higher-order rogue wave:
\begin{itemize}
 \item Select a suitable values of $s_i$, $\lambda_1$ and $\lambda_0$ to generate a certain pattern of the {\it b-positon}; Next, select suitable positon $z_0$, and then plot ideal initial pulse $q(z_0,t)$ of this  {\it b-positon};

 \item Use a frequency comb  and a wave shaper to create above ideal initial pulse $q(z_0,t)$, and then inject  it into an optical fiber;

 \item Measure the intensity of  output pulses of fiber at one or several positions $z_1, z_2, z_3,\cdots$, which are functions of $t$ denoted by $I_{1exp}(t)$, $I_{2exp}(t)$, $I_{3exp}(t)$, $\cdots$, and then compare them with one or more theoretical curves of analytical {\it b-positons}, i.e.,  $|q(z_1,t)|^2$, $|q(z_2,t)|^2$, $|q(z_3,t)|^2$, etc, in order to confirm the agreement between theory and experiment.
 \item Simulate above results of the NLS equation by a numerical way from ideal initial function, which is taken from an analytical form $q(z_0,t)$ of this {\it b-positon}, with high signal to noise ratio(SNR) (or other perturbations), to show the measurement  has high possibility in a realistic optical fiber system.
\end{itemize}
These measurements provide a good approximation of the higher-order rogue waves if we use proper parameters according to the conditions of the realistic experiment, which can be done in an optical fiber system given by  figure 1 in ref. \cite{kilberpra2014generator} or figure 3 in ref.\cite{kilberprx2015superregular}. Of course, $\lambda_1$ should be close to the $\lambda_0$ in order to get an excellent agreement between the theory and experiment.  By comparing with the observation of the first-order rogue wave  in an optical  fiber system \cite{Kibler1,Kibler2,Kibler3,Kibler4}, the main difference here is to inject  ideal (and exact) initial signals into an optical fiber in order to generate different patterns.

To illustrate this way, we provide ideal (and exact) initial pulses and theoretical output pulses  for the order-2 and order-3 {\it b-positons} according to the analytical forms in appendix with $\xi_1=-\frac{a}{2}, c>\eta_1>0$.
For a given position $z_p$ and sufficiently large $t$,  peaks in $|q^{[2]}_{\rm{{\it b\mbox{-}positon}}}(z_p,t)|^2$ and  $|q^{[3]}_{\rm{{\it b\mbox{-}positon}}}(z_p,t)|^2$  have asymptotical period $T_{asy}=\frac{\pi}{h}=\frac{\pi}{\sqrt{c^2-\eta_1^2}}$ with respect to time $t$. In the  following context, we set $c=\frac{1}{2}, \eta=\frac{2}{5}$ such at $T_{asy}=\frac{3\pi}{10}\approx 10.47$, which has been confirmed by the data in tables (1,2,3,4,5) and curves in figures (\ref{fig1x}, \ref{fig2x}, \ref{fig3x}, \ref{fig4x},\ref{fig5x}).  Because of the feature of the frequency comb system and wave shaper, the input pulse will be a periodic time series \cite{kilberprx2015superregular}, and every periodic unit is  an  ideal initial pulse in finite time length as one of  curves in figures (\ref{fig1x}, \ref{fig2x}, \ref{fig3x}, \ref{fig4x},\ref{fig5x}).  Naturally, in experiment, the output pulses are also periodic in time, and thus our theoretically predicted output pulse is just a profile of its periodic unit of time.  The asymptotical periodicity of peaks in one unit reduces the difficulty in the generation of ideal input pulses in  experiment.

  The corresponding order-2 and order-3 {\it b-positons} are plotted in figures (\ref{fig1xx},\ref{fig3xx}) in appendix. Note that we replot them again by using different values of parameters so that we can  get a higher visibility of curves in figures (\ref{fig1x}, \ref{fig2x}, \ref{fig3x}, \ref{fig4x},\ref{fig5x}), which is more helpful for works on  numerical simulation and optical observation. In figure (\ref{fig1x}), one ideal initial pulse at position $z_0=-6.80$ is plotted which  will be generated and then injected into a fiber, and one output pulse at position $z_1=0$ is plotted which denotes the predicted theoretical results of the fundamental pattern in  the central region  of an order-2 {\it b-positon} in ($z,t$)-plane. The latter will be used to compare with results of measurement in experiment which is regarded as an approximate observation of the fundamental pattern of  order-2 rogue wave.

  Similarly, for the triangular patten of the order-2 rogue wave, the fundamental pattern, triangular pattern and circular pattern of the order-3 rogue wave, we have plotted ideal initial pulses and predicted output pulses  in figures (\ref{fig2x},\ref{fig3x},\ref{fig4x},\ref{fig5x}).   The position  and  amplitude of each peak (an order-1 rogue wave), and distance of two nearest adjacent peaks in  positive axis of ideal input pulse are given in tables 1-5.

It is well known that the modulus square of an  order-$n$ rogue wave has a height $(2n+1)c^2$ and $c^2$ is the height of its asymptotic background. Moreover, an order-$n$ rogue wave can be decomposed into $\frac{n(n+1)}{2}$ uniform  peaks (an order-1 rogue wave).   According to the present values of the parameters,  the heights of the first three order rogue waves are 2.25, 6.25, 12.25.  By a close look, the output pulses in figures (\ref{fig1x}, \ref{fig3x}) are good fits of fundamental patterns of the  order-2 and order-3 rogue wave, although their heights are not coincident with these data  very well.  However, six amplitudes in figure \ref{fig5x} are not equal remarkably. There are other discrepancies in output pluses by comparing with rogue waves, which are originated from the following facts:
\begin{itemize}
\item  As long as  $\lambda_1\not=\lambda_0$,  the patterns in the central regions of an order-2 and an order-3 {\it b-positons} are not real rogue waves, and thus the above discrepancies are possible.
\item Parameters $s_i$ in {\it b-positons}, which control the decomposition of the peaks,  are not large enough.
\item In general, two peaks of the {\it b-positon} in the ($z,t$)-plane are not on a line which is parallel to the $t$ axis, so we cannot get exact amplitudes of two peaks  in one pulse by setting one value of $z$.
\end{itemize}
 In order to reduce these discrepancies, we should set $\lambda_1$ be closer to the $\lambda_0$, and set larger $s_i$, and plot more output pulses. These facts show that the observations of higher-order rogue waves are indeed difficult work. Moreover, we have to observe outputs at different positions, which also leads to difficulties for observations.

\section{Numerical simulations of the order-2 {\it b-positons}}

In a realistic optical fiber system, there are various perturbations during the propagation of the optical signals. In particular, the role of noise and hence the signal to noise ratio (SNR) is playing a key role in optical fibre communication networks. Thus, in order to take care of this important issue, it is necessary to consider errors between the theoretical results and the numerical simulations obtained  from  ideal(and exact) initial pulses with(or without) a SNR.

Figures (\ref{nscurvfundamental},\ref{nscurvtriangular}) are  simulated numerically  for fundamental and triangular patterns of the order-2 {\it b-positon} with $s_0=,a=0,c=\frac{1}{2},\lambda_1=\frac{2}{5}i$. Because of the feature of the frequency comb system and wave shaper, the input pulse will be a periodic time series \cite{kilberprx2015superregular},  figures (\ref{nscurvfundamentalextension},\ref{nscurvtriangularextension}) are simulated numerically for the periodic extension of above two cases, which can provide useful information for experiments.  The curves are not recognizable if we also put  theoretical results in figures (\ref{nscurvfundamentalextension},\ref{nscurvtriangularextension}) because the theoretical results  almost coincide completely with  the numerical simulations, so we do not add them here.

We find that there is an excellent agreement between theoretical (and exact) results and  numerical simulations when SNR $\ge$100, which shows these solutions  have strong robustness to the unavoidable noise at high SNR  in the optical fiber.
 The significant discrepancies between theoretical and numerical results  occur at the two ends of period due to the reflection of  simulation. This discrepancy is reducible by increasing the period of simulation. Therefore the fundamental and triangular patterns in order-2 {\it b-positon}  are available in realistic optical fiber system even there exists a strong noise. These results  strongly indicating the possible observation of  the higher-order rogue waves by using the central patterns of the {\it b-positons} in optical fibre systems.

 As we have pointed out  in section IV that, in order to get more accurate approximation  of the higher-order rogue waves, it is better to set $\lambda_1$ of the  {\it b-positons} to be more closer to $\lambda_0$. In other words, we could consider that the {\it b-positon}  is more closer to the rogue wave in experiment. However,  the rogue wave is extremely unstable, and thus the {\it b-positon}  with the inclusion of noise  becomes gradually but strongly unstable when $\lambda_1$ is approaching $\lambda_0$.  Therefore, it is difficult to observe the {\it b-positon} when it  is  very close to the rogue wave in experiments, not to mention the rogue wave.  The numerical simulations in figure \ref{figinstability} is clearly shows the increasing trend of the instability when $\lambda_1$ is approaching $\lambda_0$, which can be seen from the additional small  peaks appeared in  ($z,t$)-plane. Furthermore, the $X$-shape profile of the peaks are destroyed gradually in panels  by the noise from  top to bottom in figure \ref{figinstability}. Finally, the main peak of rogue wave (see bottom of the same figure) is not recognizable because it is  fully surrounded by noise peaks.

 \section{Conclusions}

 In conclusion, we have introduced  the so-called {\it b-positon} of the NLS equation, which is  obtained by taking the limit $\lambda_{j}$ $\rightarrow$ $\lambda_{1}$  in an order-$n$ breather. In other words, an order-$n$ {\it b-positon} is given by $n$ single   breathers with same height and period.  We have provided a formula expressed by the determinants and the higher-order Taylor expansion in eq.(\ref{degenerateddtqn}). It converts into an order-$n$ rogue wave by further limit $\lambda_1\rightarrow \lambda_0$. Here  $\lambda_0$ is a special eigenvalue in a single breather of the NLS  such that its period  goes to infinity, and then this breather becomes a order-1 rogue wave. We have plotted up to the order-5 {\it b-positons} in figures \ref{fig2}-\ref{fig5}.  Based on analytical formulas, we have presented a sketchy demonstration of the two limits from an order-2 breather to an order-2 rogue wave in figures \ref{fig1}-\ref{twolineatzero}.  In order to show the wide applicability of order-n {\it b-positon} in eq.(\ref{degenerateddtqn}), we also plotted the tilted propagation of  the higher-order {\it b-positons} in figures \ref{fig10}-\ref{fig30}.

 There are two main advantages of the {\it b-positon}, i.e.,  a convenient conversion to the rogue wave and the easy  controllability of the patterns in the central region in the ($z,t$)-plane.  Thus, we have suggested a new way to observe the higher-order rogue waves, namely, observe the profiles of the central region of the higher-order {\it b-positon} when $\lambda_1$ is very close to the $\lambda_0$, which can be done in an optical fiber system given by  figure 1 in ref. \cite{kilberpra2014generator} or figure 3 in ref.\cite{kilberprx2015superregular}. The ideal initial input pulse is created by a  frequency comb system and  a programable optical filter(wave shaper) according to the profile of an analytical form of the {\it b-positon} at a certain position $z_0$. We have also plotted the theoretically predicted output pulses in figures \ref{fig1x}-\ref{fig5x}, which are useful to observe the higher-order rogue waves in fiber according to suggested approach in this paper. Three patterns associated with above output pulses are plotted in figures \ref{fig1xx} and \ref{fig3xx}.

 The excellent agreements  between theoretical and numerical simulation results in figures \ref{nscurvfundamental}-\ref{nscurvtriangularextension}, and the tendency of the instability for the
 {\it b-positons} in figure \ref{figinstability} support strongly our  above new approach to observe the {\it b-positons} in a realistic optical fiber system. Our results also show the validity of the generating mechanism of a higher-order rogue wave from the double degeneracy of a multi-breather.

\textbf{Acknowledgments.}
This work is supported by the NSF of China under Grant No. 11671219,  and the K.C. Wong Magna Fund in Ningbo University.  This study is also supported by the open Fund of the State Key Laboratory of Satellite Ocean Environment Dynamics, Second Institute of Oceanography(No. SOED1708).  K. P. thanks the IFCPAR,DST, NBHM, and CSIR, Government of India, for the
financial support through major projects.

\clearpage
\section{Figures}
\subsection{Breathers and bpostions}

\begin{figure}[H]
  \centering
  \subfigure[]{\includegraphics[scale=.4]{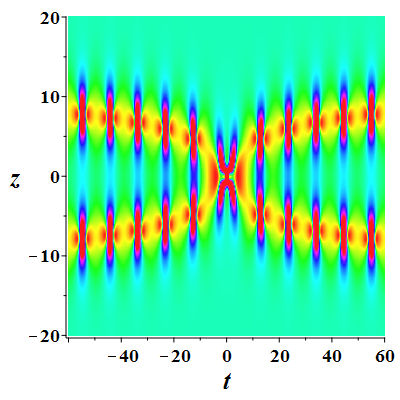}   \label{fig2:a}}
  \subfigure[]{\includegraphics[scale=.4]{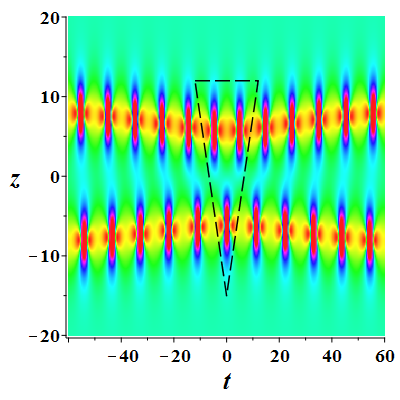}   \label{fig2:b}} \\
  \subfigure[]{\includegraphics[scale=.4]{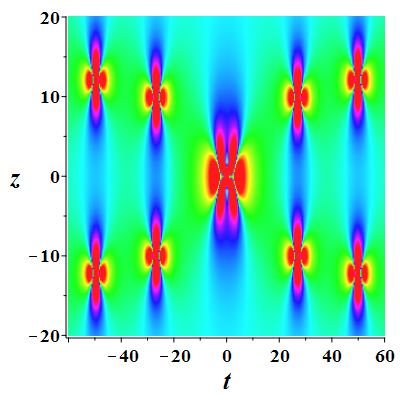}  \label{fig2:c}}
  \subfigure[]{\includegraphics[scale=.4]{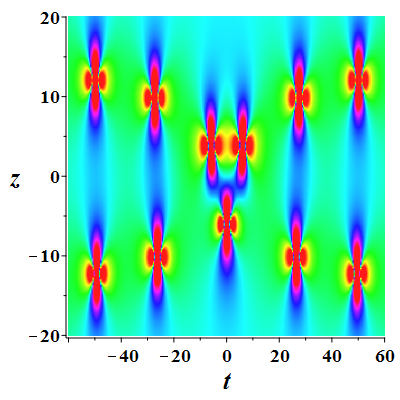}  \label{fig2:d}} \\
  \subfigure[]{\includegraphics[scale=.4]{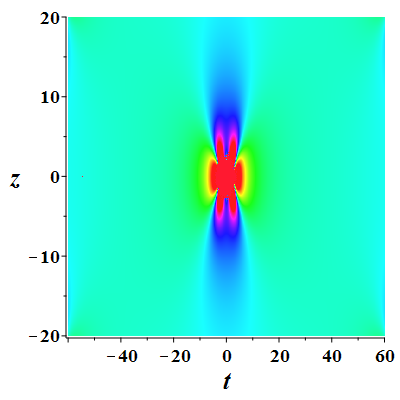}  \label{fig2:e}}
  \subfigure[]{\includegraphics[scale=.4]{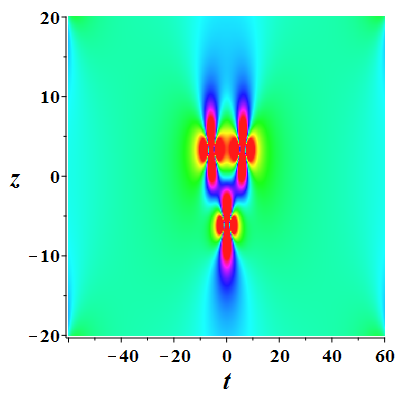}  \label{fig2:f}}
  \caption{(Color Online)
  A sketchy demonstration of the  limit  $\lambda_1 \rightarrow \lambda_0$ in an   order-2 {\it b-positon}
  $|q_{\rm{{\it b\mbox{-}positon}}}^{[2]}|^2$ (density plot) with $\lambda_0=-\frac{a}{2} + ic, a=0, c=\frac{1}{2}, s_0=0$. The central region of the left (right) column  is a fundamental (triangular) pattern. Note that the left column  is  a continuous limit of the left column in figure \ref{fig1}, but the right column is not because values of  $s_0$ in two figures are different.
  The other parameters of the {\it b-positons} respectively are
  (a) $\lambda_1=\frac{2}{5}i,s_1=0$,
  (b) $\lambda_1=\frac{2}{5}i,s_1=100$,
  (c) $\lambda_1=\frac{12}{25}i,s_1=0$,
  (d) $\lambda_1=\frac{12}{25}i,s_1=100$,
  (e) $\lambda_1=\frac{72}{145}i,s_1=0$,
  (f) $\lambda_1=\frac{72}{145}i,s_1=100$.
  }
  \label{fig2}
\end{figure}

\begin{figure}[H]
  \centering
  \subfigure[]{\includegraphics[scale=.3]{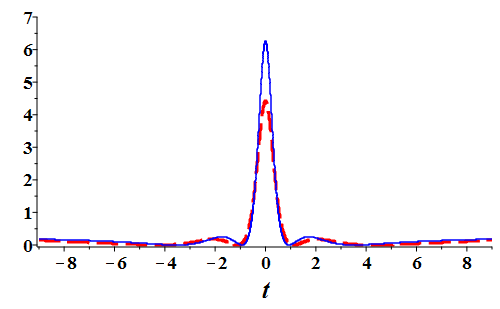}}
 \subfigure[]{\includegraphics[scale=.3]{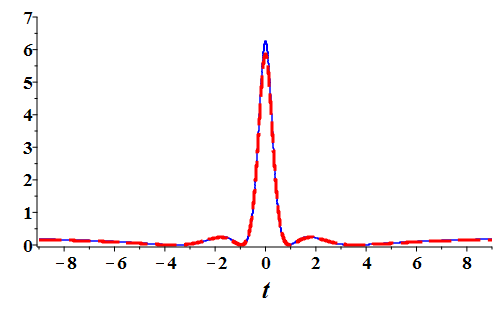}}
  \subfigure[]{\includegraphics[scale=.3]{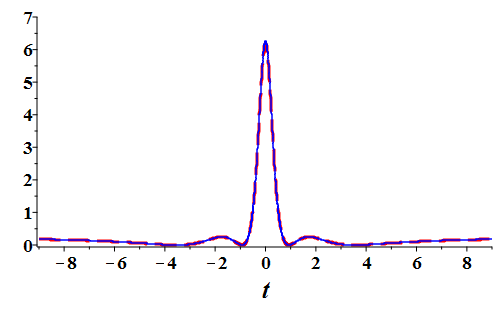}}
   \caption{(Color Online) The comparison of  an order-2 {\it b-positon} $|q_{\rm{{\it b\mbox{-}positon}}}^{[2]}|^2$(red dash line) with  an order-2 fundamental rogue wave $|q^{[2]}_{\rm rw}|^2$ (blue solid line) at $z=0$.
  The order-2 {\it b-positons} are generated through the same parameters as panels ($a,c,e$) in figure \ref{fig2} respectively. The order-2 fundamental rogue wave is generated by eq.(\ref{2rw}) with parameters $a=0,c=\frac{1}{2},s_0=s_1=0$.  The parameter $\lambda_1$ of the breathers  is  $\frac{2}{5}i$ in (a), $\frac{12}{25}i$ in (b) and $\frac{72}{145}i$ in (c).
  }
\label{twolineatzero}
\end{figure}

\begin{figure}[H]
  \centering
  \subfigure[]{\includegraphics[scale=.4]{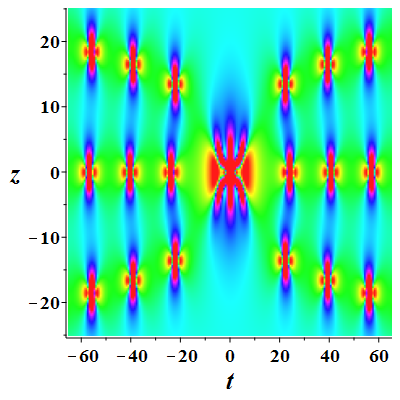}     \label{fig3:a}}
  \subfigure[]{\includegraphics[scale=.4]{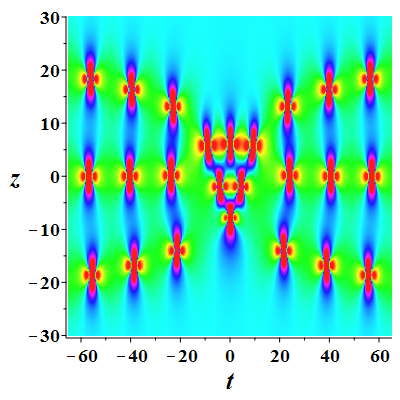}    \label{fig3:b}}\\
  \subfigure[]{\includegraphics[scale=.4]{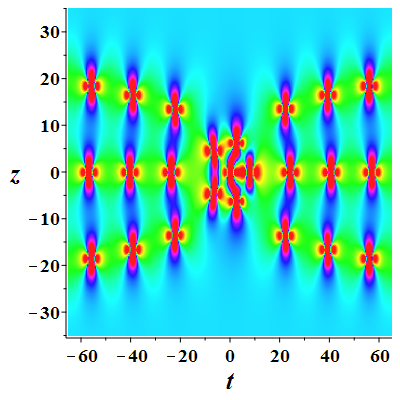}   \label{fig3:c}}
  \subfigure[]{\includegraphics[scale=.4]{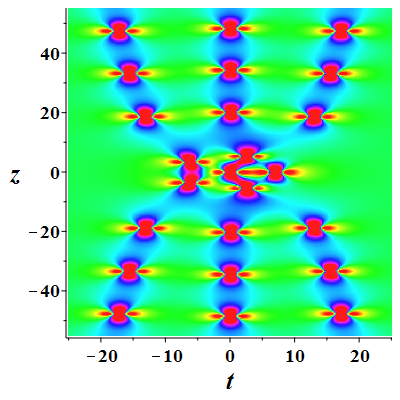}  \label{fig3:d}}
  \caption{(Color Online)  The density plots of three patterns in central region of an order-3  {\it b-positon} $|q^{[3]}_{\rm{{\it b\mbox{-}positon}}}|^2$ with $s_0=0,a=0,c=\frac{1}{2}$.
  (a) The fundamental pattern with $s_1=0,s_2=0,\lambda_1=\frac{6}{13} i$,
  (b) the triangular pattern with $s_1=50,s_2=0,\lambda_1=\frac{6}{13} i$,
  (c) the circular pattern with $s_1=0,s_2=500,\lambda_1=\frac{6}{13} i$,
  (d) the circular pattern with different directions with $s_1=0,s_2=500,\lambda_1=\frac{13}{24} i$.  }
  \label{fig3}
\end{figure}


\begin{figure}[H]
  \centering
  \subfigure[]{\includegraphics[scale=.4]{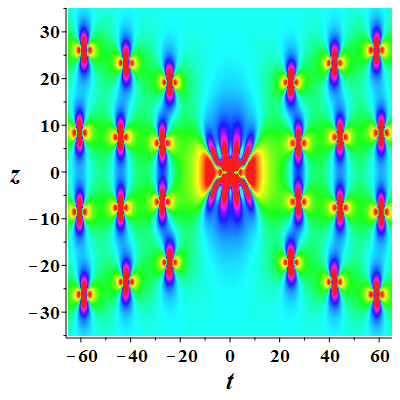}   \label{fig4:a}}
  \subfigure[]{\includegraphics[scale=.4]{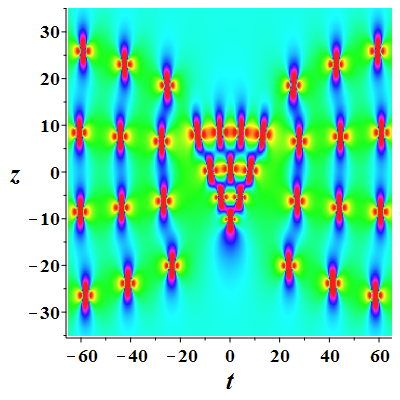}   \label{fig4:b}}\\
  \subfigure[]{\includegraphics[scale=.4]{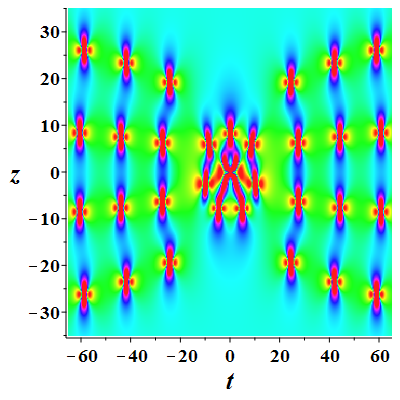} \label{fig4:c}}
  \subfigure[]{\includegraphics[scale=.4]{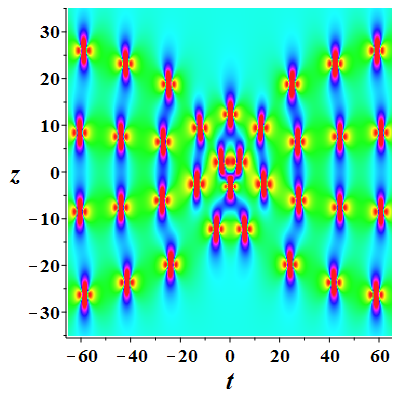}  \label{fig4:d}}
  \caption{(Color Online)  The density plots of four patterns in central region of an order-4  {\it b-positon} $|q^{[4]}_{\rm{{\it b\mbox{-}positon}}}|^2$ with
  $a=0,c=\frac{1}{2},s_0=s_2=0,\lambda_1=\frac{6}{13} i$ .
  (a) The fundamental pattern with $s_1=0,s_3=0$,
  (b) the triangular pattern with $s_1=50,s_3=0$,
  (c) the circular pattern with an inner fundamental pattern when $s_1=0,s_3=5\times10^3$,
  (d) the circular pattern with an inner triangular pattern when $s_1=20,s_3=5\times10^4$.
  }
  \label{fig4}
\end{figure}


\begin{figure}[H]
  \centering
  \subfigure[]{\includegraphics[scale=.4]{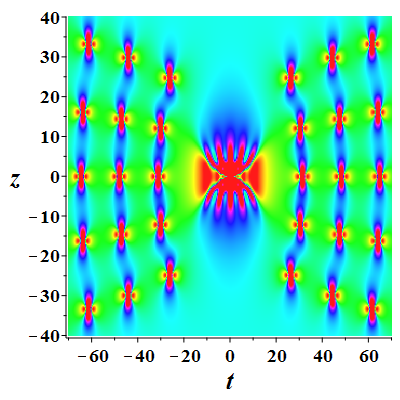}    \label{fig5:a}}
  \subfigure[]{\includegraphics[scale=.4]{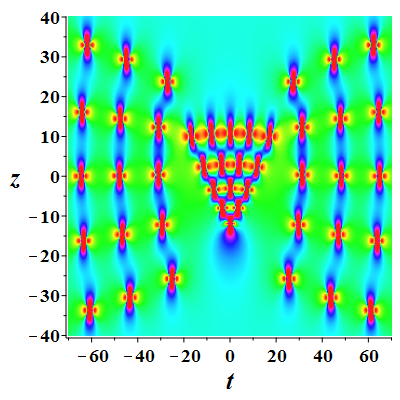}   \label{fig5:b}}\\
  \subfigure[]{\includegraphics[scale=.4]{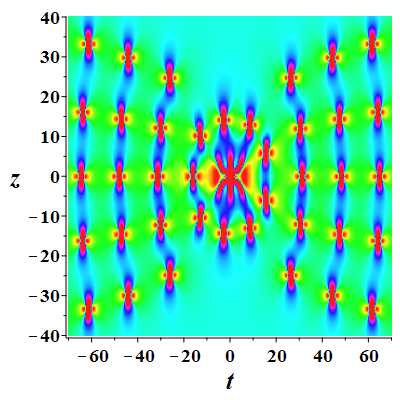}     \label{fig5:c}}
  \subfigure[]{\includegraphics[scale=.4]{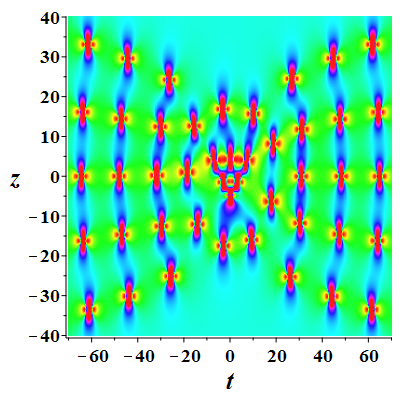}  \label{fig5:d}} \\
  \subfigure[]{\includegraphics[scale=.4]{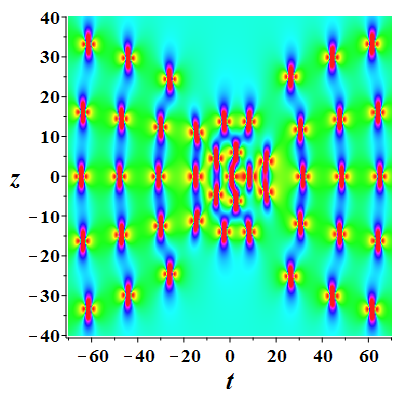} \label{fig5:e}}
  \subfigure[]{\includegraphics[scale=.4]{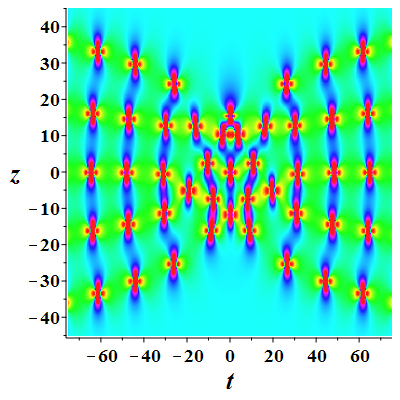}     \label{fig5:f}}
  \caption{(Color Online)  The density plots of six patterns in central region of an order-5  {\it b-positon} $|q^{[5]}_{\rm{{\it b\mbox{-}positon}}}|^2$ with $a=0,c=\frac{1}{2},s_0=0,\lambda_1=\frac{6}{13} i$.
  (a) The fundamental pattern with $s_1=0,s_2=0,s_3=0,s_4=0$,
  (b) the triangular pattern with $s_1=50,s_2=0,s_3=0,s_4=0$,
  (c) the circular pattern with an inner fundamental pattern when $s_1=0,s_2=0,s_3=0,s_4=5\times10^5$,
  (d) the circular pattern with an inner triangular pattern when $s_1=20,s_2=0,s_3=0,s_4=2\times10^6$,
  (e) the circular pattern with inner decomposed peaks when $s_1=0,s_2=500,s_3=0,s_4=5\times10^5$,
  (f) two-ring pattern with $s_1=0,s_2=0,s_3=5\times10^4,s_4=0$.
  }
  \label{fig5}
\end{figure}

\clearpage
\subsection{Ideal initial input pulses and predicted output pulses}

\begin{figure}[H]
  \centering
  \subfigure[]{\includegraphics[scale=.4]{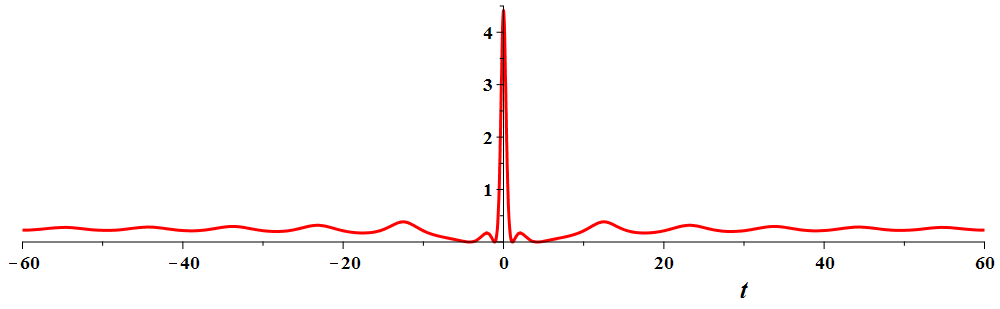}  \label{fig1x:a0}} \\
  \subfigure[]{\includegraphics[scale=.4]{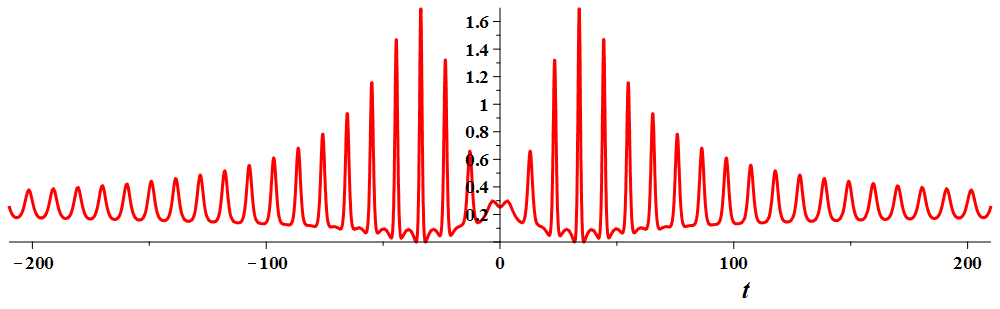}  \label{fig1x:b0}}
  \caption{(Color Online)  Pulses of the fundamental pattern of an  order-2  {\it b-positon} with $ c=\frac{1}{2} , \lambda_1=\frac{2}{5} i ,a=s_0=s_1=0 $.   (a) Output pulse (theoretically predicted ) at $z_1=0$, the amplitude is   $4.41$ at $t=0$.    (b) Ideal initial input pulse at $z_0=- 6.80$. Note that the amplitude is associated with the main peak in figures \ref{fig1xx}(a,b).}
  \label{fig1x}
\end{figure}

\begin{table}[b]
\centering
\caption{Data of peaks in figure \ref{fig1x}}
\begin{tabular}{|c|c|c|c|c|c|c|c|c|c|c|}
  \hline
  $t_i$ & 12.89 & 23.38 &  33.87 &  44.36 &  54.85 &  65.33 &  75.82 &  86.30 &  96.78 & 107.25   \\ \hline
  $h$ & 0.66 &  1.32 &  1.69 &  1.47 &  1.16 &  0.93 &  0.78 &  0.68 &  0.61 &  0.56   \\ \hline
  $\Delta t_i$ & $\diagup$ & 10.48 &  10.49 & 10.49 &  10.49 &  10.48 &  10.48 &  10.48 &  10.48 &  10.48    \\
  \hline
\end{tabular}
{\\ Notes: $\Delta t_i= t_i-t_{i-1}$, $t_i$ denotes the time of a peak, $h$ denotes the amplitude of a peak.}
\end{table}

\clearpage

\begin{figure}[H]
  \centering
  \subfigure[]{\includegraphics[scale=.4]{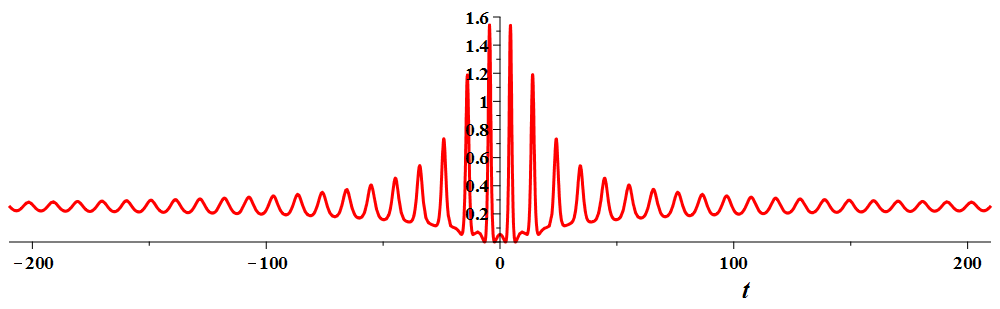}  \label{fig2x:a0}} \\
  \subfigure[]{\includegraphics[scale=.4]{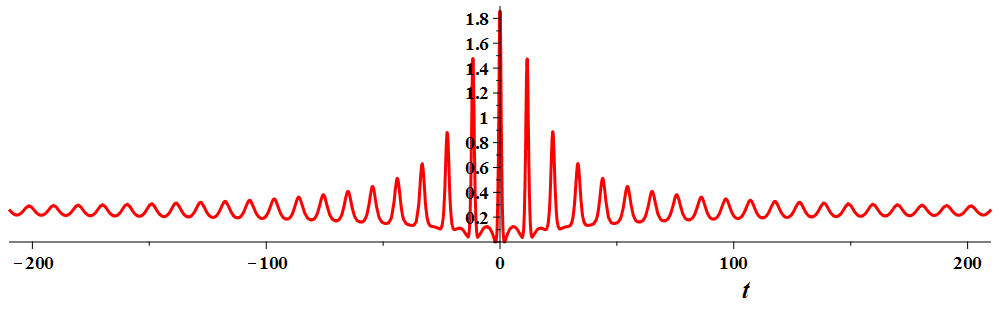}  \label{fig2x:b0}} \\
  \subfigure[]{\includegraphics[scale=.4]{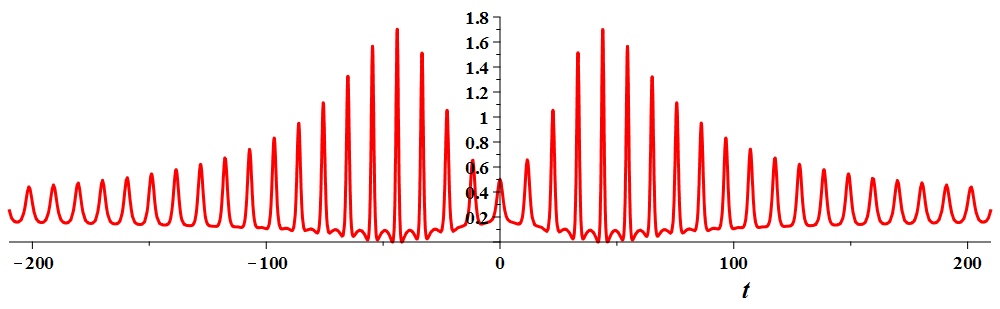}  \label{fig2x:c0}}
  \caption{(Color Online)   Pulses of the triangular  pattern of an order-2  {\it b-positon}
   with $ c=\frac{1}{2} , \lambda_1=\frac{2}{5} i, a=s_0=0, s_1=50$.
  (a) Output pulse-2 (theoretically predicted) at $z_2=4.42$, the amplitude is  $1.54$ at $t=\pm 4.47$.
  (b) Output pulse-1 (theoretically predicted) at $z_1=-4.81$, the amplitude is  $1.86$ at $t=0$.
  (c) Ideal initial input pulse  at $z_0= -7.43$. Note that three amplitudes are associated with the triplets in figures \ref{fig1xx}(c,d) around the coordinate origin.}
  \label{fig2x}
\end{figure}
\begin{table}[b]
\centering
\caption{Data of peaks in figure \ref{fig2x}}
\begin{tabular}{|c|c|c|c|c|c|c|c|c|c|c|}
  \hline
  $t_i$ & 11.69 & 22.65 & 33.36 & 43.96 & 54.53   & 65.06 &75.58 & 86.09& 96.59 & 107.08 \\ \hline
  $h$ & 0.66 & 1.053 & 1.513 & 1.701  & 1.564   &  1.33 & 1.11 & 0.95 & 0.83 & 0.74 \\ \hline
  $\Delta t_i$ & $\diagup$ & 10.96 & 10.70 & 10.61 & 10.56  & 10.53 &10.52  & 10.51 & 10.50 & 10.50 \\
  \hline
\end{tabular}
{\\ Notes: $\Delta t_i= t_i-t_{i-1}$, $t_i$ denotes the time of a peak, $h$ denotes the amplitude of a peak.}
\end{table}
\clearpage

\begin{figure}[H]
\centering
  \subfigure[]{\includegraphics[scale=.4]{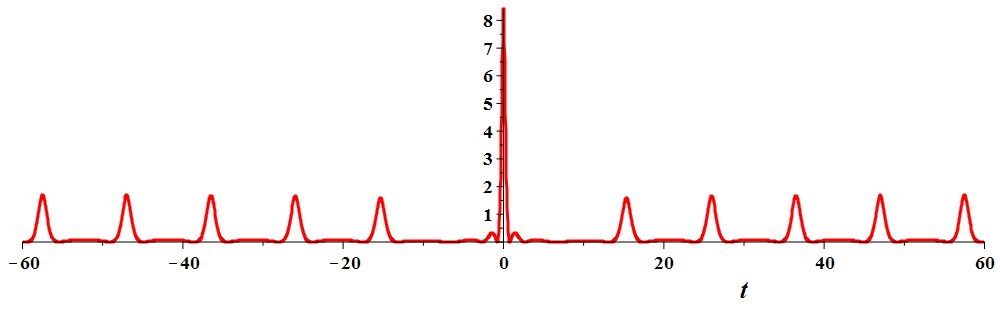}  \label{fig3x:a0}} \\
  \subfigure[]{\includegraphics[scale=.4]{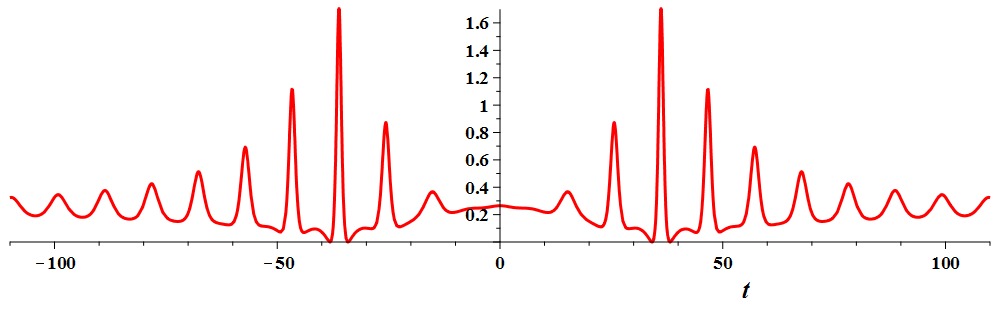}  \label{fig3x:b0}}
  \caption{(Color Online) Pulses of the fundamental pattern of an  order-3  {\it b-positon}  with $ c=\frac{1}{2} , \lambda_1=\frac{2}{5} i ,a=s_0=s_1=s_2=0 $.  (a) Output pulse (theoretically predicted) at $z_1=0$, the amplitude is $8.41$ at $t=0$. (b) Ideal initial input   pulse at $z_0= -12.4$.  Note that the amplitude is associated with the main peak in figures \ref{fig3xx}(a,b). }
  \label{fig3x}
\end{figure}
\begin{table}[b]
\centering
\caption{Data of peaks in figure \ref{fig3x}}
\begin{tabular}{|c|c|c|c|c|c|c|c|c|c|c|}
  \hline
  $t_i$ & 15.16 & 25.63 & 36.15 &  46.67 &  57.18 &  67.69 &  78.20 &  88.70 &  99.19 &  109.69  \\ \hline
  $h$ & 0.37 &  0.87 &  1.70 &  1.12 &  0.69 &  0.51 &  0.43 &  0.38 &  0.35 &  0.33  \\ \hline
  $\Delta t_i$ & $\diagup$ &  10.48 &  10.51 &  10.52 &  10.51 &  10.51 &  10.50 &  10.50 &  10.50 &  10.49  \\
  \hline
\end{tabular}
{\\ Notes: $\Delta t_i= t_i-t_{i-1}$, $t_i$ denotes the time of a peak, $h$ denotes the amplitude of a peak.}
\end{table}
\clearpage

\begin{figure}[H]
\centering
  \subfigure[]{\includegraphics[scale=.4]{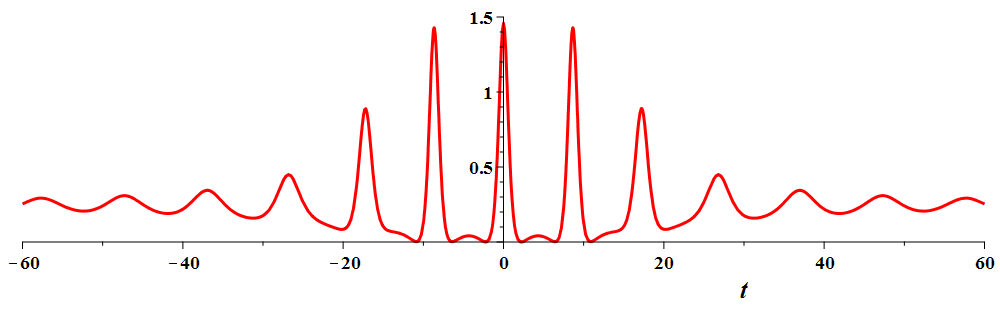}  \label{fig4x:a0}} \\
  \subfigure[]{\includegraphics[scale=.4]{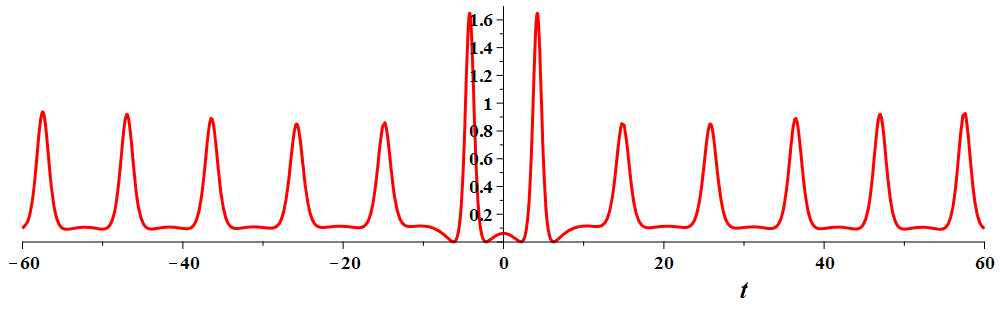}  \label{fig4x:b0}} \\
  \subfigure[]{\includegraphics[scale=.4]{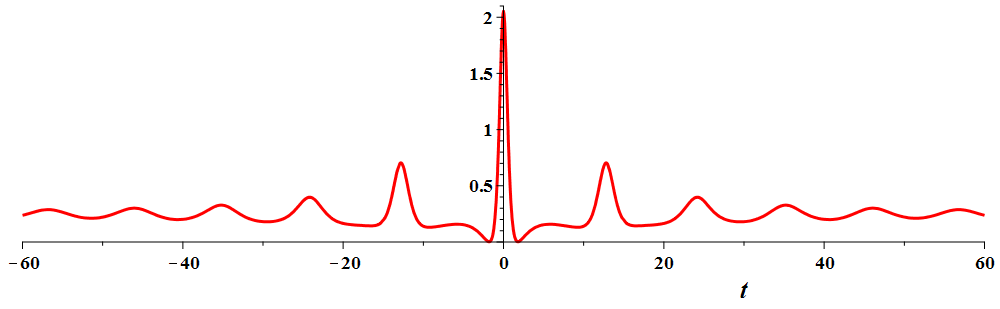}  \label{fig4x:c0}} \\
  \subfigure[]{\includegraphics[scale=.4]{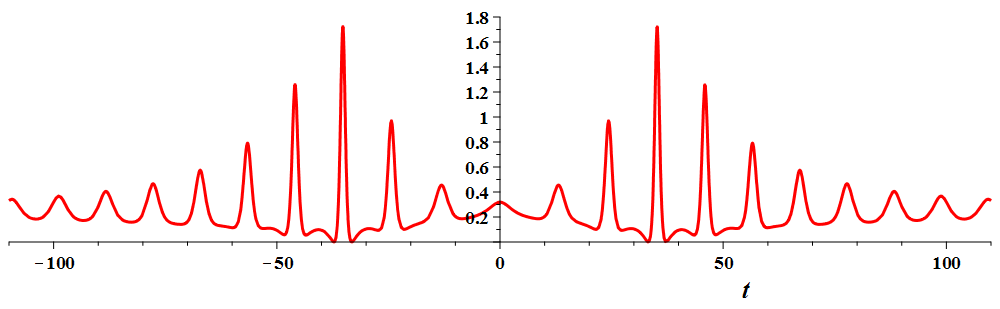}  \label{fig4x:d0}}
  \caption{(Color Online) Pulses of the triangular  pattern of an  order-3  {\it b-positon}  with $ c=\frac{1}{2} , \lambda_1=\frac{2}{5} i ,a=s_0=s_2=0, s_1=50 $. (a) Output pulse-3 (theoretically predicted) at $z_3=8$, the amplitudes  are $1.43$ at
  $t=\pm 8.65 $ and  $1.46$ at $t=0$.
  (b) Output pulse-2 (theoretically predicted) at $z_2=-1.28$, the amplitudes are  $1.65$ at $t=\pm 4.21$.
  (c) Output pulse-1 (theoretically predicted) at($z_1=-8$),  the amplitude $2.05$ at $t=0$.
  (d) Ideal initial input pulse at $z_0= -12.8$. Note that the six amplitudes are associated with the six peaks in triangular  figures \ref{fig3xx}(c,d).}
  \label{fig4x}
\end{figure}

\begin{table}
\centering
\caption{Data of peaks in figure \ref{fig4x}}
\begin{tabular}{|c|c|c|c|c|c|c|c|c|c|c|}
  \hline
  $t_i$ &13.10 & 24.34 & 35.20 & 45.92 & 56.56 & 67.16 & 77.73 & 88.29 & 98.83 & 109.35     \\ \hline
  $h$ & 0.46 & 0.97 & 1.72 & 1.26 & 0.79 & 0.57 & 0.47 & 0.41 & 0.37 & 0.34    \\ \hline
  $\Delta t_i$ & $\diagup$ &  11.24 & 10.87 & 10.72 & 10.64 & 10.60 & 10.57 & 10.55 & 10.54 & 10.53   \\
  \hline
\end{tabular}
{\\ Notes:$\Delta t_i= t_i-t_{i-1}$, $t_i$ denotes the time of a peak, $h$ denotes the amplitude of a peak.}
\end{table}

\begin{table}
\centering
\caption{Data of peaks in figure \ref{fig5x}}
\begin{tabular}{|c|c|c|c|c|c|c|c|c|c|c|}
  \hline
  $t_i$ &2.12 & 14.49 & 25.41 & 36.04 & 46.60 & 57.14 & 67.66 & 78.17 & 88.68 & 99.18  \\ \hline
  $h$ & 0.30 & 0.40 & 0.93 & 1.71 & 1.08 & 0.67 & 0.50 & 0.42 & 0.37 & 0.34      \\ \hline
  $\Delta t_i$ & $\diagup$ &  12.33 & 10.93 & 10.62 & 10.56 & 10.54 & 10.52 & 10.51 & 10.51 & 10.50     \\
  \hline
\end{tabular}
{\\ Notes: $\Delta t_i= t_i-t_{i-1}$, $t_i$ denotes the time of a peak, $h$ denotes the amplitude of a peak.}
\end{table}

\clearpage

\begin{figure}[H]
\centering
  \subfigure[]{\includegraphics[scale=.4]{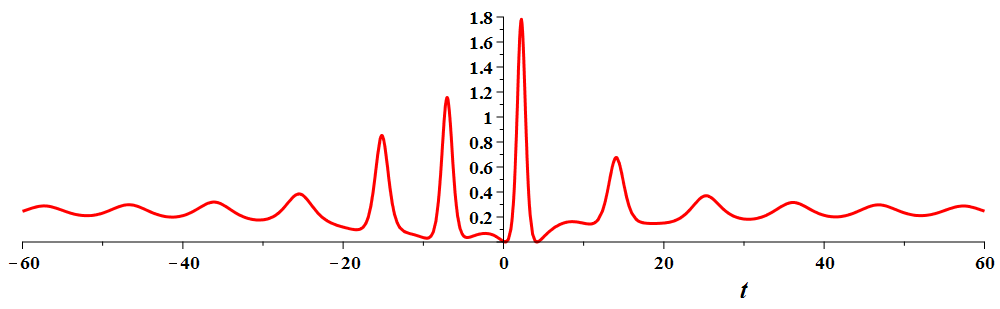}  \label{fig4x:a0}} \\
  \subfigure[]{\includegraphics[scale=.4]{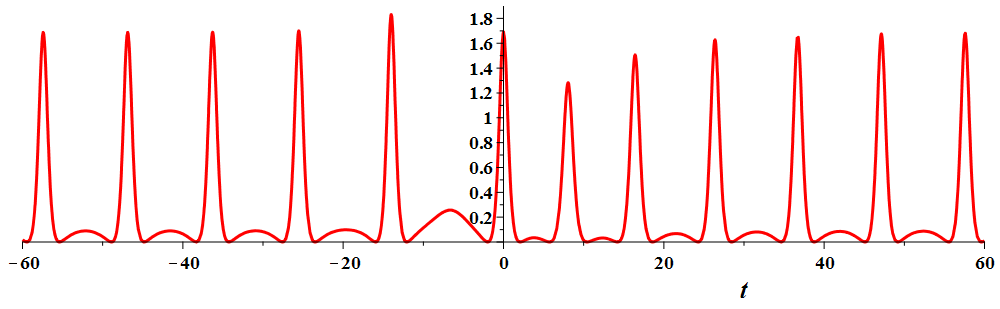}  \label{fig4x:b0}} \\
  \subfigure[]{\includegraphics[scale=.4]{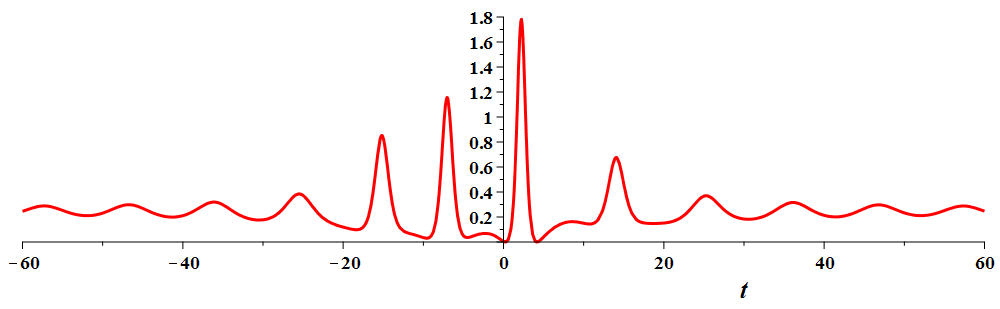}  \label{fig4x:c0}} \\
  \subfigure[]{\includegraphics[scale=.4]{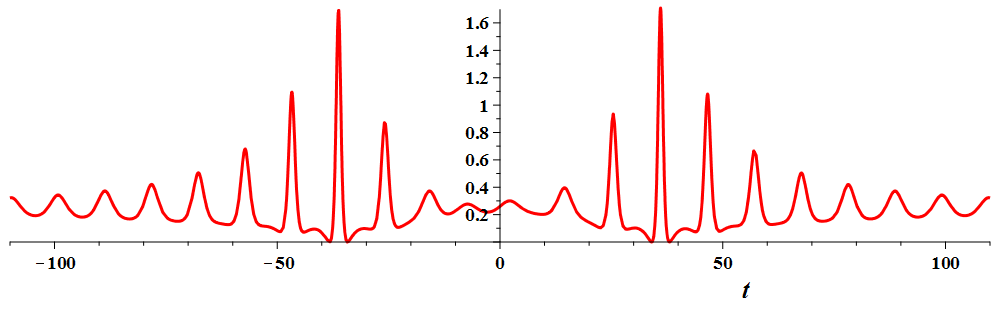}  \label{fig4x:d0}}
  \caption{(Color Online)  Pulses of the circular pattern of an  order-3  {\it b-positon}  with $ c=\frac{1}{2} , \lambda_1=\frac{2}{5} i ,a=s_0=s_1=0, s_2=500 $.   (a) Output pulse-3 (theoretically predicted) at $z_3=7$, the amplitudes  are $1.16$ at $t=-7.03$ and  $1.78$ at
  $t=2.22$).
  (a) Output pulse-2 (theoretically predicted) at $z_2=0$, the amplitudes  are $1.69$ at $t=0$ and  $1.28$ at
  $t=8.06$.
  (c) Output pulse-1 (theoretically predicted) at $z_1=-7$ is the same as the result of $z_3$.
  (d) Ideal initial input pulse at $z_0=-12.35$. Note that the six amplitudes are associated with six peaks in circle of figures \ref{fig3xx}(e,f). }
  \label{fig5x}
\end{figure}


\clearpage

\section{Appendix}

\subsection{Rogue waves}
\subsubsection{ Order-1 Rogue wave }
Taking the limit $\lambda_1\rightarrow \lambda_0$, the single breather yields the  firsr-order rogue wave
\begin{equation}\label{1rw}
q^{[1]}_{\rm rw}=\left( \frac{4 (1+ iT)} {  X^2 + T^2 +  1 } -1 \right) c { e}^{i\rho},
\end{equation}
with
$$ X=2c \left( t-2az+\Re{(s_0)} \right),  T=2c\left(2cz+\Im{(s_0)}\right). $$

\subsubsection{ Order-2 Rogue wave }
Taking the limit $\lambda_1\rightarrow \lambda_0$ and setting $\Im{(s_0)}=0$ , order-2 {\it b-positon} yields  an order-2 rogue wave
\begin{equation}\label{2rw}
q^{[2]}_{\rm rw}= \left( 1 -  \frac{12 \left(F_{\rm rw}^{[2]} + i G_{\rm rw}^{[2]} \right)}{ H_{\rm rw}^{[2]} }   \right)  c  { e}^{i\rho}.
\end{equation}
Here
$$F_{\rm rw}^{[2]}=5\,{T}^{4}+6\,{T}^{2}{X}^{2}+{X}^{4}+18\,{T}^{2}+6\,{X}^{2}-3 + 24\,{c}^{2} \left( T\Re{(s_1)}-X\Im{(s_1)} \right)  $$

$$
G_{\rm rw}^{[2]}=12\,{c}^{2} \left( {T}^{2}-{X}^{2}-1 \right) \Re{(s_1)} -24\,{c}^{2} T X \Im{(s_1)} +
 \left( {T}^{4}+2\,{T}^{2}{X}^{2}+{X}^{4}+2\,{T}^{2}-6\,{X}^{2}-15  \right) T
$$

$$H_{\rm rw}^{[2]}=H_{{\rm rw} 1}^{[2]}+H_{{\rm rw} 2}^{[2]}$$

$$
H_{{\rm rw} 1}^{[2]}= 144 c^4 {|s_1|^2} + 24{c^2}  \left( \left( {T^2}-3{X^2}+9 \right) T \Re{(s_1)} + \left({X^2} -3{T^2}-3 \right) X \Im{(s_1)}   \right)
$$

$$H_{{\rm rw} 2}^{[2]}= \left( {T^2}+{X^2} \right)^{3}+3\left( 3{T^2}-{X^2} \right)^{2}+99{T^2}+27{X^2} + 9 $$

$$ X=2c \left( t-2az+\Re{(s_0)} \right),  T=2c\left(2cz+\Im{(s_0)}  \right)= 4 c^2 z $$

$\Re{( \cdot )}$ denotes the real part, $\Im{( \cdot )}$ denotes the imaginary part.

\newpage

\subsection{ {\it b-positons} (order-2 and order-3) }
\subsubsection{ Order-2 Bpositons }
 Set $ \lambda_1=\xi_1+i\eta_1, \xi_1=-\frac{a}{2}, S_0=s_0+s_1\epsilon $  and $n=2$ in $q^{[n]}_{\rm{{\it b\mbox{-}positon}}}$, an order-2 {\it b-positon} is given by
 $$
q^{[2]}_{\rm {\it b-positon}}(z,t; a,c,\eta_1;s_0,s_1)= \left( \frac { F^{[2]}+4\,i\eta_1\,h G^{[2]}} {H^{[2]}}   \right) 8 c^2  { e}^{i\rho}.
$$
Here
$\rho = a t + (2 c^2 - a^2) z ,  h=h(\lambda_1)=\sqrt{c^2-\eta_1^2},$

$s_0=s_{0R}+i s_{0I}, s_1=s_{1R}+i s_{1I}$

$F^{[2]}=
F_1 \cos \left( 2\,X_{{0}} \right) +
F_2 \cosh \left( 2\,T_{{0}} \right) +
F_3 \cosh \left( T_{{0}} \right) \cos \left( X_{{0}} \right) +
F_4 \cosh \left( T_{{0}} \right) \sin \left( X_{{0}} \right) +
F_5 \sinh \left( T_{{0}} \right) \cos \left( X_{{0}}  \right) +
F_6
$

$
$

$
F_1= -c{\eta_1}^{4}, \quad
F_2= c \left( {c}^{4}-8\,{\eta_1}^{2}{h}^{2} \right) ,
$

$
F_3= 4\,\eta_1\,{h}^{2} \left( {c}^{2}+2\,{\eta_1}^{2} \right) , \quad
F_4= 8\,{\eta_1}^{2}h\omega_{{2}}X_{{1}} ,
$

$
F_5= -8\,{\eta_1}^{2}h \omega_{{2}}T_{{1}}, \quad
F_6= c {h}^{2} \left( 8\,{\eta_1}^{2} F_7 + {c}^{2}-7\,{\eta_1}^{2} \right)
$

$
F_7=  {\eta_1}^{2} \left( {X}^{2}+{s_{0I}}^{2}  \right) + {h}^{2} \left(   |s_1|^{2}
 {h}^{2}+2\,\eta_1\, \left( X{s_{1I}}-s_{0I}\,{s_{1R}} \right)  \right) +4\,T_{{2}}
$

\[\]

$G^{[2]}=
c\omega_{{2}} \sinh \left( 2\,T_{{0}} \right)
-4\,{\eta_1}^{2}hT_{{1}} \cosh \left( T_{{0}} \right) \cos \left( X_{{0}} \right) +
2\,{\eta_1}^{3} \cos \left( X_{{0}} \right) \sinh \left( T_{{0}} \right) +
4\,{\eta_1}^{2}hX_{{1}}\sinh \left( T_{{0}} \right) \sin \left( X_{{0}} \right) +
4\,c\eta_1\,hT_{{1}}
$

 \[\]

$H^{[2]}=
32\,{c}^{3}\eta_1\,h H_1 +
64\,{c}^{2}{\eta_1}^{2}{h}^{2} H_2   +
8\,{c}^{2}{h}^{2} H_3
+8\,{c}^{6}\cosh \left( 2\,T_{{0}} \right)
-8\,{c}^{2}{\eta_1}^{4}\cos \left( 2\,X_{{0}} \right)
$

$
$

$H_1=
2\,\eta_1\,T_{{1}}\sinh \left( T_{{0}}  \right) \cos \left( X_{{0}} \right)
-2\,\eta_1\,X_{{1}}\cosh \left( T_{{0}} \right) \sin \left( X_{{0}} \right)
-h \cosh \left( T_{{0}} \right) \cos \left( X_{{0}} \right)
$

$H_2=
\eta_1\,X_{{2}}+4\,T_{{2}}
$

$H_3=
8\,{\eta_1}^{2}  |s_1|^{2}  {h}^{4}-16
\,{\eta_1}^{3} s_{0I}\,{s_{1R}}\,{h}^{2}+8\,{\eta_1}^{4}{s_{0I}}^{2}+{c}^{2}+{
\eta_1}^{2}$

and

$ X_0= 2 h X ,  X_1={s_{1I}} h^2 + X\eta_1,  X_2=X\left(2{s_{1I}} h^2+X\eta_1\right), X=t-2az+{s_{0R}}$

$ T_0= 2 h (2 \eta_1 z + s_{0I}) , T_1 = 2 \omega_2  z + \omega_3 , T_2= \omega_2  z \left(\omega_2  z + \omega_3\right) $

$\omega_2=c^2-2\eta_1^2,  \omega_3  = h^2{s_{1R}} - \eta_1 s_{0I} $\\

\newpage

\subsubsection{Order-3 Bpositons with Fundamental pattern}
Set $ \lambda_1=\xi_1+i\eta_1, \xi_1=-\frac{a}{2},s_i=0$  and $n=3$ in $q^{[n]}_{\rm{{\it b\mbox{-}positon}}}$, an order-3 {\it b-positon} is given by

$$
q_{\rm{{\it b\mbox{-}positon}}}^{[3]}(z,t;a,c, \eta_1)=\left(   \frac{F^{[3]}+i\mu\sqrt{1-\mu^2}G^{[3]}}{H^{[3]}}  \right) c e^{i\rho}
$$

{\footnotesize
\begin{eqnarray*}
H^{[3]}=\cosh^{3}(T) + \mu\, H_1 \cosh^{2}(T) +  2\,T\mu  H_2 \sinh(T)\cosh(T) + H_3 \cosh(T)
-2\,T H_4 \sinh(T) +
 {\mu}^{9}\cos^{3}(X) +
 \mu\, H_5 \cos(X)
 -2\,X{\mu}^{3} H_6 \sin(X)
\end{eqnarray*}

\begin{eqnarray*}
H_1= 2\,X{\mu}^{2
} \left( 5\,{\mu}^{2}-4 \right) \sin(X) - \left( 6\,{\mu
}^{4}-8\,{\mu}^{2}+3-2\,{X}^{2}{\mu}^{4}+2\, \left( 2\,{\mu}^{2}-1
 \right) ^{2}{T}^{2} \right) \cos(X)
\end{eqnarray*}

\begin{eqnarray*}
H_2=  \left( 12\,{\mu
}^{4}-14\,{\mu}^{2}+3 \right) \cos(X) -2\,X{\mu}^{2}
 \left( 2\,{\mu}^{2}-1 \right) \sin(X)
\end{eqnarray*}

\begin{eqnarray*}
H_3=2\,X{\mu}^{6} \left( {\mu}^{2}-2 \right) \cos(X) \sin(X)
-{\mu}^{4} H_{31}   \cos^{2}(X)
+ H_{32}
\end{eqnarray*}

\begin{eqnarray*}
H_{31}= 3\,{\mu}^{4}-8\,{\mu}^{2}+6-2\,{X}^{2}{\mu}^{4}+2\, \left( 2\,{\mu}^{2}-1
 \right) ^{2}{T}^{2}
\end{eqnarray*}

\begin{eqnarray*}
H_{32}=  \left(  \left( 2\,{\mu}^{2}-1 \right) ^{2}{T}^{2}+{X}^{2}{\mu}^{4} \right) ^{2}+{\mu}^{4} \left( 5\,{\mu}^{4}-12\,{\mu}^{2}+6
 \right) {X}^{2}+ \left( 64\,{\mu}^{8}-128\,{\mu}^{6}+88\,{\mu}^{4}-24
\,{\mu}^{2}+3 \right) {T}^{2}
\end{eqnarray*}

\begin{eqnarray*}
H_4 = {\mu}^{4} \left( 4\,{\mu}^{2}-3 \right) \cos^{2}(X)
+2\,X{\mu}^{6} \left( 2\,{\mu}^{2}-1 \right)  \cos(X) \sin(X) + \left( {\mu}^{4} \left( 4\,{\mu}^{4}-6\,{\mu}^{2}+3
 \right) {X}^{2}  + \left( 8\,{\mu}^{4}-8\,{\mu}^{2}+1 \right)  \left( 2 \,{\mu}^{2}-1 \right) ^{2}{T}^{2} \right)
\end{eqnarray*}

\begin{eqnarray*}
H_5= \left(  \left( 2\,{\mu}^{2}-1 \right) ^{2}{T}^{2}+{X}^{2}{\mu}^{4}
 \right)^{2}-3\,{X}^{2}{\mu}^{8}+ \left( -24\,{\mu}^{8}+24\,{\mu}^{6}
+18\,{\mu}^{4}-20\,{\mu}^{2}+3 \right) {T}^{2}
\end{eqnarray*}

\begin{eqnarray*}
H_6={X}^{2}{\mu}^{6}+ \left( 2\,{\mu}^{2}-1
 \right)  \left( 6\,{\mu}^{4}-3\,{\mu}^{2}-2 \right) {T}^{2}
\end{eqnarray*}

\clearpage

{\footnotesize
\begin{eqnarray*}
F^{[3]}= F_1 \cosh(T)^{3} -\mu\, F_2 \cosh(T) ^{2} +2\,T\mu\, F_3 \cosh(T) \sinh(T)
+ F_4 \cosh(T)
+2\,T
 \left( 2\,{\mu}^{2}-1 \right)  F_5 \sinh(T) - {\mu}^{9} \cos^{3}(X)-\mu\, F_6 \cos(X)
+2\,X{\mu}^{3} F_7 \sin(X)
\end{eqnarray*}

\begin{eqnarray*}
F_1=- \left( 2\,{\mu}^{2}-1 \right)  \left( 16\,{\mu}^{4}-16\,{\mu}^{2}+1  \right)
\end{eqnarray*}
\begin{eqnarray*}
F_2= F_{21} \cos(X) +2\,X{\mu}^{2} F_{22} \sin(X)
\end{eqnarray*}
\begin{eqnarray*}
F_{21}=  \left( 8\,{\mu}^{4}-8\,{\mu}^{2}+1 \right)  \left( 2\,{X}^{2}
{\mu}^{4}-2\, \left( 2\,{\mu}^{2}-1 \right) ^{2}{T}^{2} \right) +2\,{
\mu}^{2} \left( 8\,{\mu}^{6}-24\,{\mu}^{4}+21\,{\mu}^{2}-4 \right) -3
\end{eqnarray*}
\begin{eqnarray*}
F_{22}=  8\,{\mu}^{6}-24\,{\mu}^{4}+21\,{\mu}^{2}-4
\end{eqnarray*}

\begin{eqnarray*}
F_3=  2\,X{\mu
}^{2} \left( 2\,{\mu}^{2}-1 \right)  \left( 8\,{\mu}^{4}-8\,{\mu}^{2}+
1 \right) \sin(X) - \left( 32\,{\mu}^{8}-80\,{\mu}^{6}+
68\,{\mu}^{4}-22\,{\mu}^{2}+3 \right) \cos(X)
\end{eqnarray*}
\begin{eqnarray*}
F_4= {\mu}^{4} \left( F_{41} \cos(X) +  F_{42}  \sin(X) \right)  \cos(X) + F_{43}
\end{eqnarray*}
\begin{eqnarray*}
F_{41}=  -2\, \left( 2\,{\mu}^{2}-1 \right)  \left( {X}^{2}{\mu
}^{4}- \left( 2\,{\mu}^{2}-1 \right) ^{2}{T}^{2} \right) +6\,{\mu}^{6}
-3\,{\mu}^{4}+4\,{\mu}^{2}-6
\end{eqnarray*}
\begin{eqnarray*}
F_{42}= 2\,X{\mu}^{2} \left( 6\,{\mu}^{4}-3\,{\mu}^{2}-2 \right)
\end{eqnarray*}
\begin{eqnarray*}
F_{43}= - \left( 2\,{\mu}^{2}-1 \right)
 \left(  \left( 2\,{\mu}^{2}-1 \right) ^{2}{T}^{2}+{X}^{2}{\mu}^{4}
 \right) ^{2}+{\mu}^{4} \left( 6\,{\mu}^{6}-3\,{\mu}^{4}-8\,{\mu}^{2}+
6 \right) {X}^{2}-3\, \left( 2\,{\mu}^{2}-1 \right)  \left( 8\,{\mu}^{
4}-8\,{\mu}^{2}+1 \right) {T}^{2}
\end{eqnarray*}

\begin{eqnarray*}
F_5= {\mu}^{4} \left(  \left( 8\,{
\mu}^{4}-4\,{\mu}^{2}-3 \right) \cos(X) +2\,X{\mu}^{2}
 \left( 2\,{\mu}^{2}-1 \right) \sin(X)  \right) \cos
 \left( X \right) -{\mu}^{4} \left( 4\,{\mu}^{4}-2\,{\mu}^{2}-3
 \right) {X}^{2}+ \left( 2\,{\mu}^{2}-1 \right) ^{2}{T}^{2}
\end{eqnarray*}
\begin{eqnarray*}
F_6=  -3\,{X}^{2}{\mu}^{8}+ \left( 104\,{\mu}^{8}-232
\,{\mu}^{6}+178\,{\mu}^{4}-52\,{\mu}^{2}+3 \right) {T}^{2}+ \left(
 \left( 2\,{\mu}^{2}-1 \right) ^{2}{T}^{2}+{X}^{2}{\mu}^{4} \right) ^{2}
\end{eqnarray*}
\begin{eqnarray*}
F_7=  {X}^{2}{\mu}^{6
}+ \left( 2\,{\mu}^{2}-1 \right)  \left( 6\,{\mu}^{4}-3\,{\mu}^{2}-2
 \right) {T}^{2}
\end{eqnarray*}

\begin{eqnarray*}
G^{[3]}= G_1 \sinh(T)  \cosh(T) ^{2}
+ G_2 \cosh(T)^{2}
+8\,\mu\, G_3 \cosh(T) \sinh(T)
+4\,T G_4 \cosh(T)
 + G_5 \sinh(T)
+8\,T\mu\, G_6
\end{eqnarray*}

\begin{eqnarray*}
G_1=  2\, \left( 4\,{\mu}^{2}-1 \right)  \left( 4\,{\mu}^{2}-3 \right)
\end{eqnarray*}
\begin{eqnarray*}
G_2=  16\,T \mu\, \left( 2\,{\mu}^{2}-1 \right) ^{2} \left(  \left( {\mu}^{2}-1
 \right) \cos(X) -X{\mu}^{2}\sin(X)
 \right)
\end{eqnarray*}
\begin{eqnarray*}
G_3=  X{\mu}^{2} \left( {\mu}^{2}-1 \right)  \left( 2\,{\mu}^{2}-3 \right)
\sin(X) + G_{31} \cos(X)
\end{eqnarray*}
\begin{eqnarray*}
G_{31}=  {\mu}^{4} \left( 2\,{\mu}^{2}-1
 \right) {X}^{2}- \left( 2\,{\mu}^{2}-1 \right) ^{3}{T}^{2}+{\mu}^{2}
 \left( {\mu}^{2}-1 \right)  \left( 2\,{\mu}^{2}-3 \right)
\end{eqnarray*}

\begin{eqnarray*}
G_4=  \left( 2\,{\mu}^{2}-1 \right) ^{2}{T}^{2}+{\mu}
^{4} \left( 4\,{\mu}^{4}-2\,{\mu}^{2}-1 \right) {X}^{2}+16\,{\mu}^{2}
 \left( {\mu}^{2}-1 \right) +3-{\mu}^{4} \left(  \left( 8\,{\mu}^{4}-4
\,{\mu}^{2}-1 \right) \cos(X)
+2\,X{\mu}^{2} \left( 2\,{
\mu}^{2}-1 \right) \sin(X)  \right) \cos \left( X
 \right)
\end{eqnarray*}

\begin{eqnarray*}
G_5= 2 G_{51} -2\,{\mu}^{4} \left(  G_{52}\cos(X) +6\,X{\mu}^{4}\sin(X)  \right)\cos(X)
\end{eqnarray*}

\begin{eqnarray*}
G_{51}=   \left(  \left( 2 \,{\mu}^{2}-1 \right) ^{2}{T}^{2}+{X}^{2}{\mu}^{4} \right) ^{2}
-3\,{X}^{2}{\mu}^{8}  - \left( 4\,{\mu}^{2}-1 \right)  \left( 4\,{\mu}^{2}-3  \right) {T}^{2}
\end{eqnarray*}

\begin{eqnarray*}
G_{52}=   2\, \left( 2\,{\mu}^{2}-1 \right) ^{2}{T}^{2}-{\mu}^{4} \left( 2\,{X}^{2}-3 \right)
\end{eqnarray*}

\begin{eqnarray*}
G_6=    \left(  \left( 2\,{\mu}^{2}-1 \right) ^{3}{T}^{2}+{\mu}^{4}
 \left( 2\,{\mu}^{2}-1 \right) {X}^{2}+{\mu}^{2} \left( {\mu}^{2}-1
 \right)  \left( 2\,{\mu}^{2}-3 \right)  \right) \cos \left( X
 \right) +X{\mu}^{2} \left( {\mu}^{2}-1 \right)  \left( 2\,{\mu}^{2}-3
 \right) \sin(X)
\end{eqnarray*}

\[
\mu=\frac{\eta_1}{c}, X=2h(t-2az),  T = 4 h \eta_1 z,  h=c\sqrt{1-\mu^2}
\]

}

\clearpage

\subsection{The tilted propagation of the {\it b-positons} (order-2 and order-3)}

\begin{figure}[H]
  \centering
  \subfigure[]{\includegraphics[scale=.4]{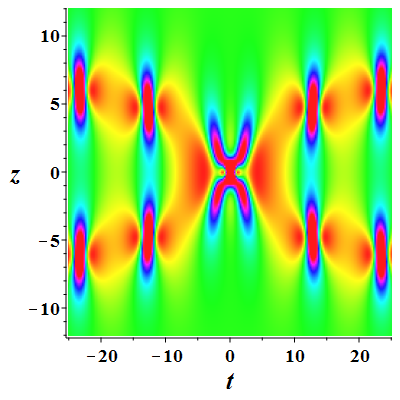}  \label{fig10:a0}}
  \subfigure[]{\includegraphics[scale=.4]{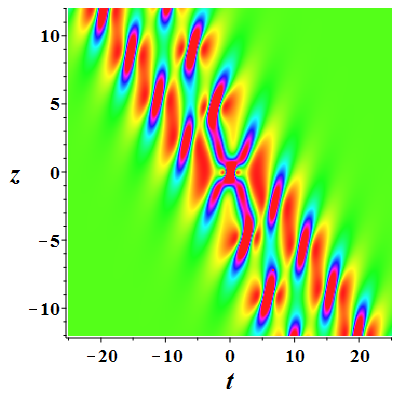}  \label{fig10:b0}} \\
  \subfigure[]{\includegraphics[scale=.4]{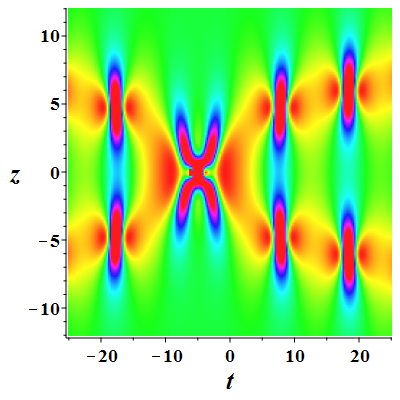}  \label{fig10:c0}}
  \subfigure[]{\includegraphics[scale=.4]{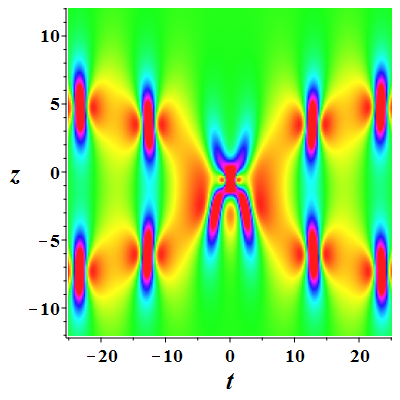}    \label{fig10:d0}}
  \caption{(Color Online)  The evolution of an order-2
  {\it b-positon} $|q^{[2]}_{\rm{{\it b\mbox{-}positon}}}|^2$ (density plot) on ($z, t$)-plane with  $c=\frac{1}{2} ,s_1=0, \lambda_1=\frac{2}{5} i$.
  (a) The fundamental pattern with $s_0=0,a=0$;
  (b) the titled propagation with a minor rotation when $s_0=0,a=\frac{1}{2}$;
  (c) the fundamental pattern with $s_0=5,a=0$, which has been shifted along  negative direction of $t$ axis about $\Re(s_0)$ unit by comparing with (a);
  (d) the fundamental pattern with $s_0=i,a=0$, which has been shifted along  negative direction of
    $z$ axis with deformation of the profile in central region  about $\Im(s_0)$ unit by comparing with (a). These shifts are
    originated from contributions of $\Re(s_0)$ or $\Im(s_0)$. }
  \label{fig10}
\end{figure}

\begin{figure}
  \centering
  \subfigure[]{\includegraphics[scale=.4]{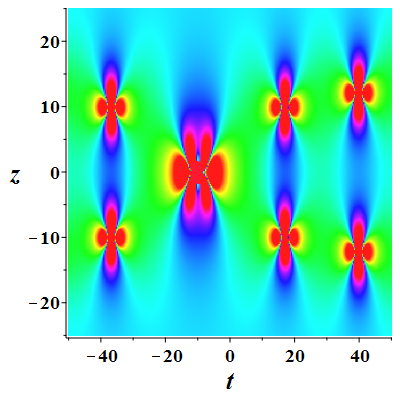}  \label{fig20:c0}}
  \subfigure[]{\includegraphics[scale=.4]{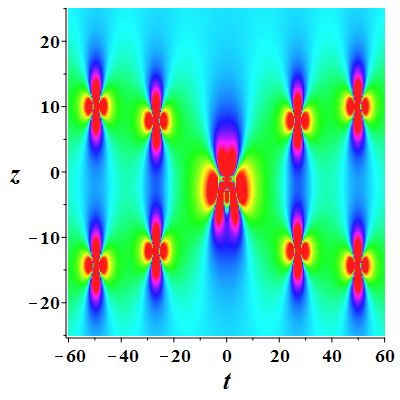}  \label{fig20:d0}}
  \caption{(Color Online)   The evolution of an order-2   {\it b-positon} $|q^{[2]}_{\rm{{\it b\mbox{-}positon}}}|^2$ (density plot) on ($z, t$)-plane with  $a=0,c=\frac{1}{2}, \lambda_1=\frac{12}{25} i$.
  (a) The fundamental pattern with $s_0=10,s_1=0$,
  (b) the triangular pattern with $s_0=2i,s_1=0$.
  By comparing with the two pictures in last row of figure \ref{fig10}, there exist a more larger shift of the
  central profile, and also a significantly increasing of the distance of two peaks along the time axis. This effects reflect
  the contribution of the non-zero value of $s_0$. }
  \label{fig20}
\end{figure}
\begin{figure}
  \centering
  \subfigure[]{\includegraphics[scale=.4]{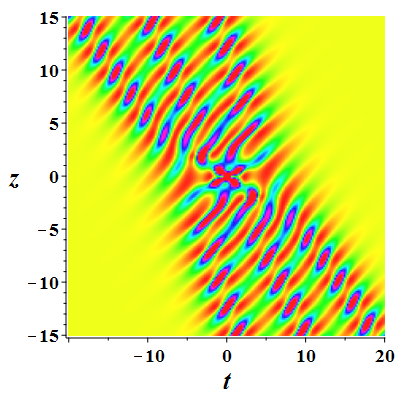} \label{fig30:a}}
  \subfigure[]{\includegraphics[scale=.4]{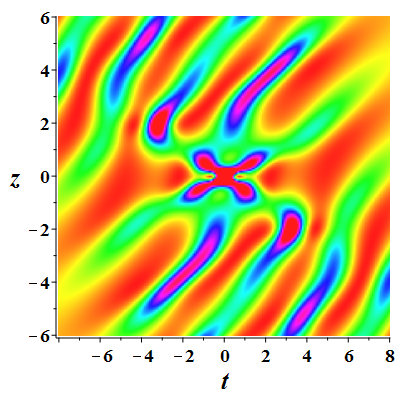}     \label{fig30:b}}\\
  \subfigure[]{\includegraphics[scale=.4]{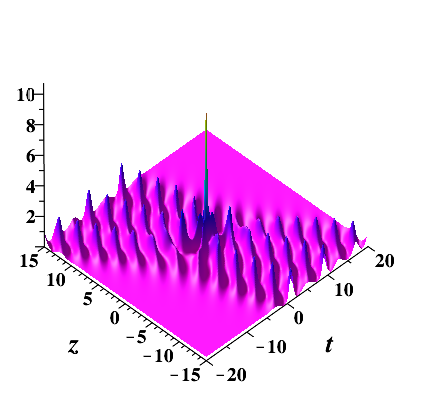} \label{fig30:c}}
  \subfigure[]{\includegraphics[scale=.4]{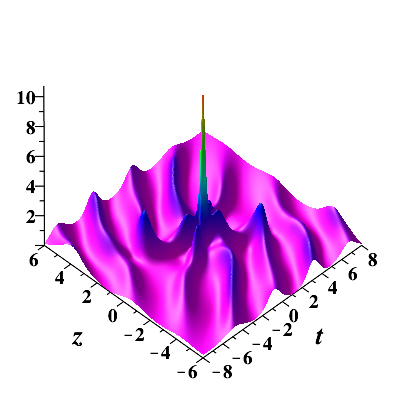}     \label{fig30:d}}
  \caption{(Color Online)  The tilted propagation of an order-3 {\it b-positon} $|q^{[3]}_{\rm{{\it b\mbox{-}positon}}}|^2$   on ($z, t$)-plane with $a=1,c=\frac{1}{2},s_0=s_1=s_2=0,\lambda_1=\frac{6}{13} i$.
  (a) The density plot,
  (b) the central profile of (a),
  (c) three dimensional profile of $|q^{[3]}_{\rm{{\it b\mbox{-}positon}}}|^2$,
  (d) the central profile of (c).}
  \label{fig30}
\end{figure}

\clearpage

\subsection{  {\it B-positons}(order-2 and order-3) for experiments}

\begin{figure}[H]
  \centering
  \subfigure[]{\includegraphics[scale=.4]{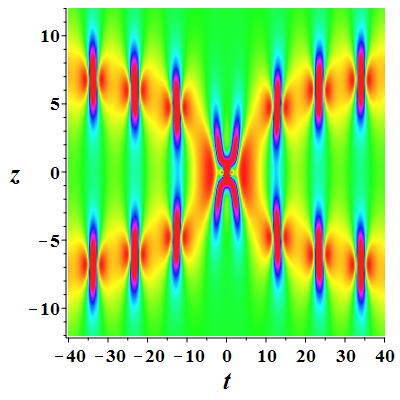}  \label{fig1xx:a0}}
  \subfigure[]{\includegraphics[scale=.4]{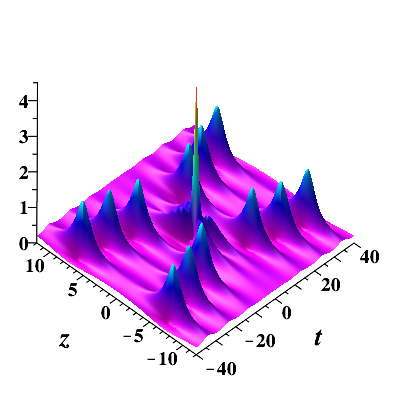}  \label{fig1xx:b0}} \\
  \subfigure[]{\includegraphics[scale=.4]{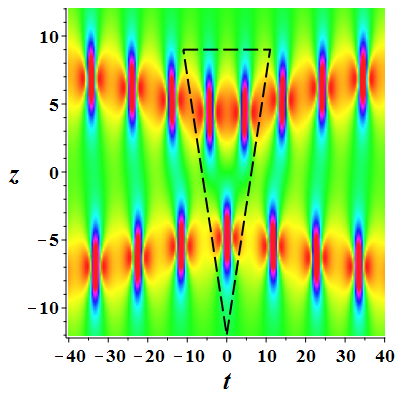}  \label{fig1xx:c0}}
  \subfigure[]{\includegraphics[scale=.4]{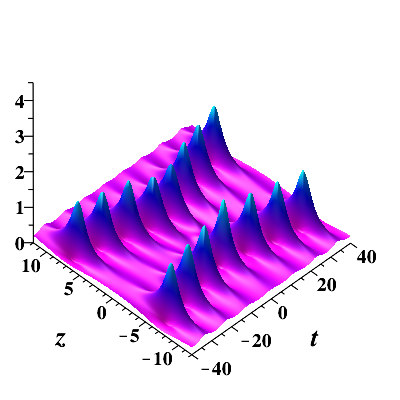}  \label{fig1xx:d0}}
  \caption{(Color Online)  Two patterns  of an order-2 {\it b-positon} $|q^{[2]}_{\rm{{\it b\mbox{-}positon}}}|^2$ with $ c=\frac{1}{2} , \lambda_1=\frac{2}{5} i,   a=s_0=0$.
  (a) The fundamental pattern with $s_1=0$ (density plot),
  (b) the fundamental pattern with $s_1=0$  (three dimensional profile),
  (c) the triangular pattern with $s_1=50$  (density plot),
  (d) the triangular pattern with  $s_1=50$ (three dimensional profile).
  The ideal initial input and output(theoretically predicted) pulses are given in figures \ref{fig1x} and \ref{fig2x}. }
  \label{fig1xx}
\end{figure}

\begin{figure}
  \centering
  \subfigure[]{\includegraphics[scale=.4]{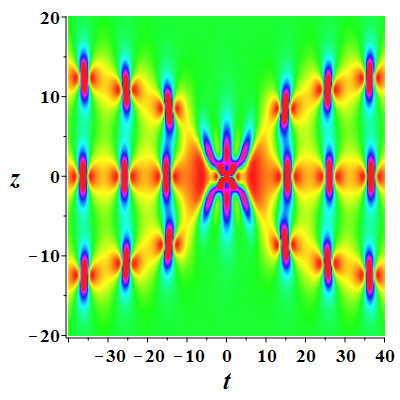}  \label{fig3xx:a0}}
  \subfigure[]{\includegraphics[scale=.4]{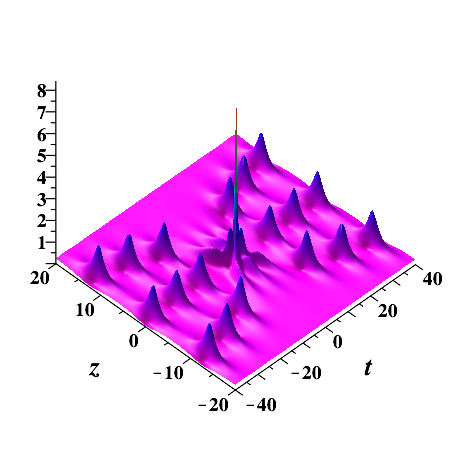}  \label{fig3xx:b0}} \\
  \subfigure[]{\includegraphics[scale=.4]{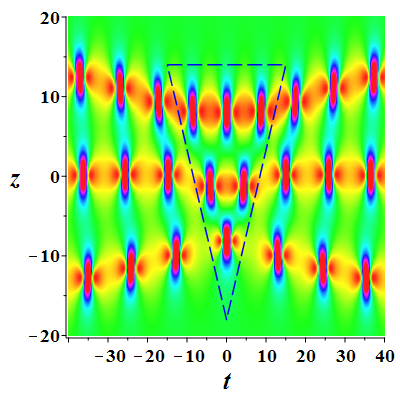}  \label{fig3xx:a0}}
  \subfigure[]{\includegraphics[scale=.4]{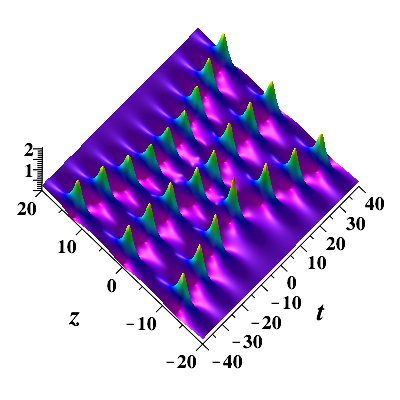}  \label{fig3xx:b0}} \\
  \subfigure[]{\includegraphics[scale=.4]{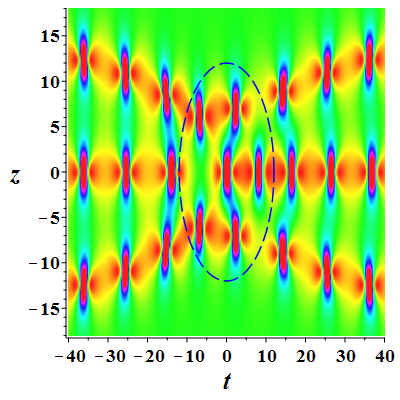}  \label{fig3xx:a0}}
  \subfigure[]{\includegraphics[scale=.4]{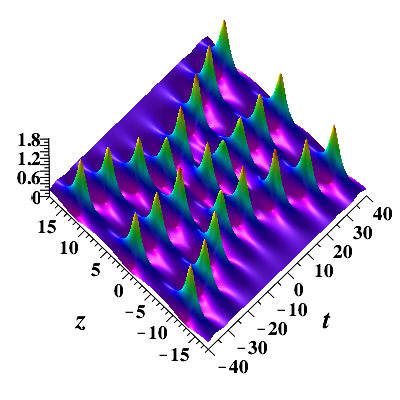}  \label{fig3xx:b0}}
  \caption{(Color Online)
  Three patterns  of an order-3 {\it b-positon} $|q^{[3]}_{\rm{{\it b\mbox{-}positon}}}|^2$ with $ c=\frac{1}{2} , \lambda_1=\frac{2}{5} i ,a=s_0=0 $.
  (a)  The fundamental pattern with $s_1=s_2=0$ (density plot),
  (b) the fundamental pattern with $s_1=s_2=0$  (three dimensional profile),
  (c) the triangular pattern with $s_1=50,s_2=0$  (density plot),
  (d) the triangular pattern with $s_1=50,s_2=0$  (three dimensional profile),
  (e) the circular pattern with $s_1=0,s_2=500$  (density plot),
  (f) the circular pattern with $s_1=0,s_2=500$  (three dimensional profile).
   The ideal initial input and output(theoretically predicted) pulses are given in figures \ref{fig3x}, \ref{fig4x} and \ref{fig5x}.}
  \label{fig3xx}
\end{figure}

\clearpage

\subsection{Numerical simulations for order-2 {\it b-positons}}

The numerical code for the NLS equation is given in ref.\cite{agrawalbook5ed}(see its appendix B). In following figures, the first column denotes input signals, others denote output signals.

\[\]

\begin{figure}[H]
  \centering
  \includegraphics[scale=.4]{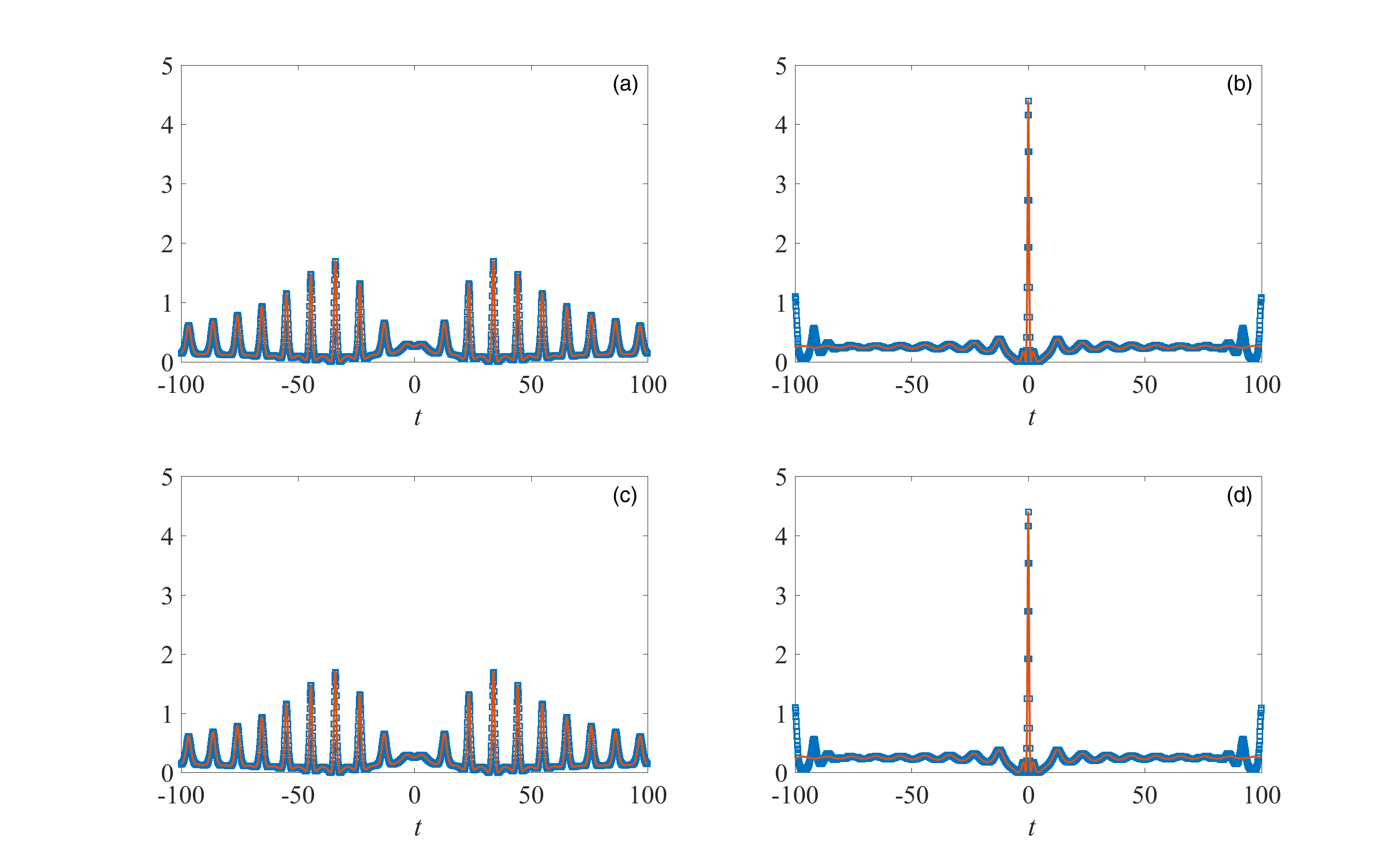}\\
   \caption{(Color Online)  Numerical simulation  of the fundamental pattern in central region of an order-2  {\it b-positon} $|q^{[2]}_{\rm{{\it b\mbox{-}positon}}}|^2$ with $s_0=s_1=0,a=0,c=\frac{1}{2},\lambda_1=\frac{2}{5}i$. Red line  denotes the theoretical (and exact) result, and square point denotes the result of numerical simulation.  The significant discrepancies occur
    at the two ends of period due to the reflection of simulation.
  (a) The initial signal of fundamental pattern without noise at $z=-6.80$,
  (b) The observe signal of fundamental pattern without noise at $z=0$,
  (c) The initial signal of fundamental pattern with noise at $z=-6.80$ (SNR=100),
  (d) The observe signal of fundamental pattern with noise at $z=0$ (SNR=100).
  }
  \label{nscurvfundamental}
\end{figure}

\clearpage

\[\]
\begin{figure}[H]
  \centering
  \includegraphics[scale=.4]{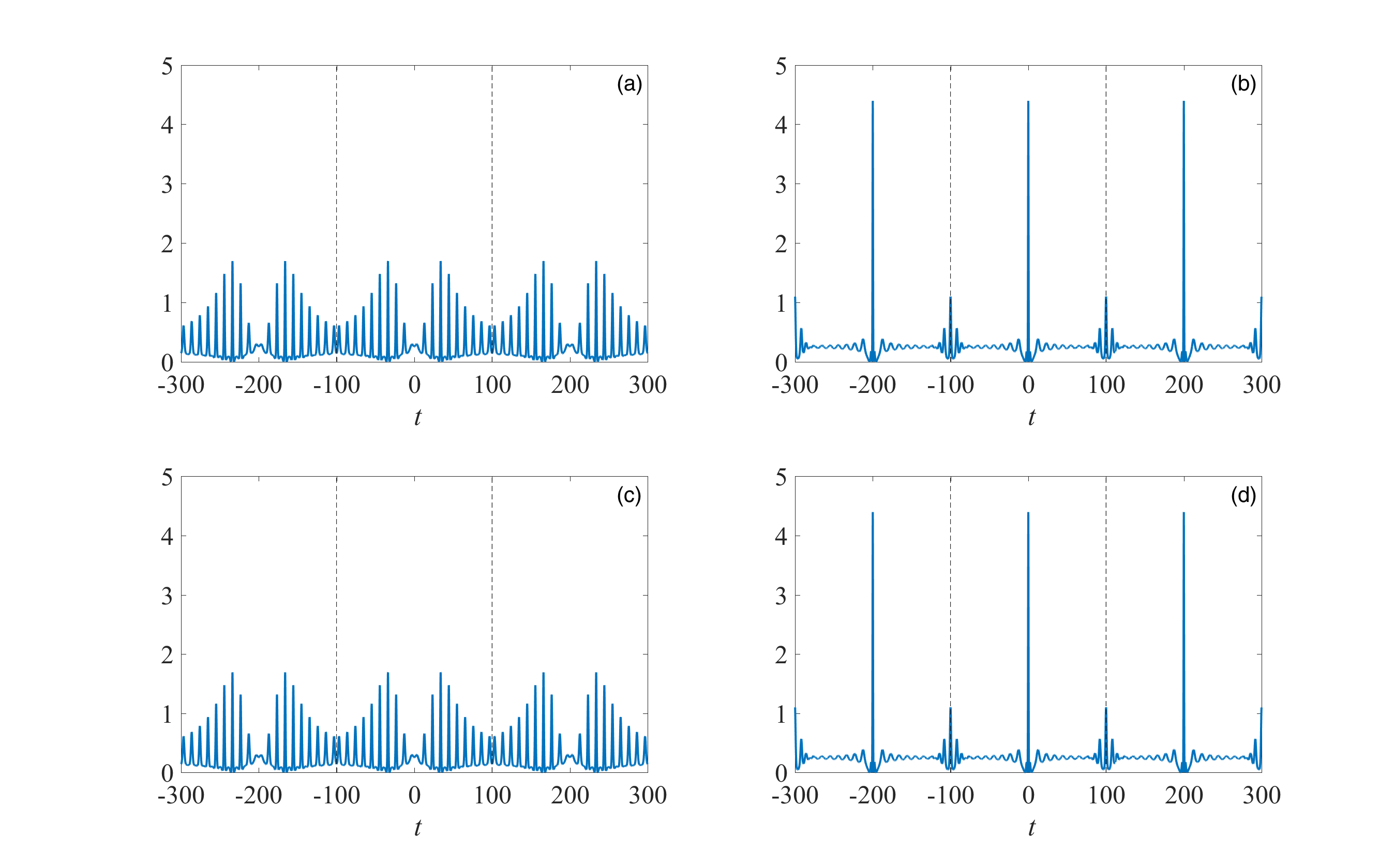}\\
   \caption{(Color Online) Numerical simulation of three periods from periodic extension of figure \ref{nscurvfundamental}. In order to get recognizable curve, theoretical results are not added.
  (a) The initial signal of fundamental pattern without noise at $z=-6.80$,
  (b) The observe signal of fundamental pattern without noise at $z=0$,
  (c) The initial signal of fundamental pattern with noise at $z=-6.80$ (SNR=100),
  (d) The observe signal of fundamental pattern with noise at $z=0$ (SNR=100).
  }
  \label{nscurvfundamentalextension}
\end{figure}

\clearpage

\[\]

\begin{figure}[H]
  \centering
  \includegraphics[scale=.4]{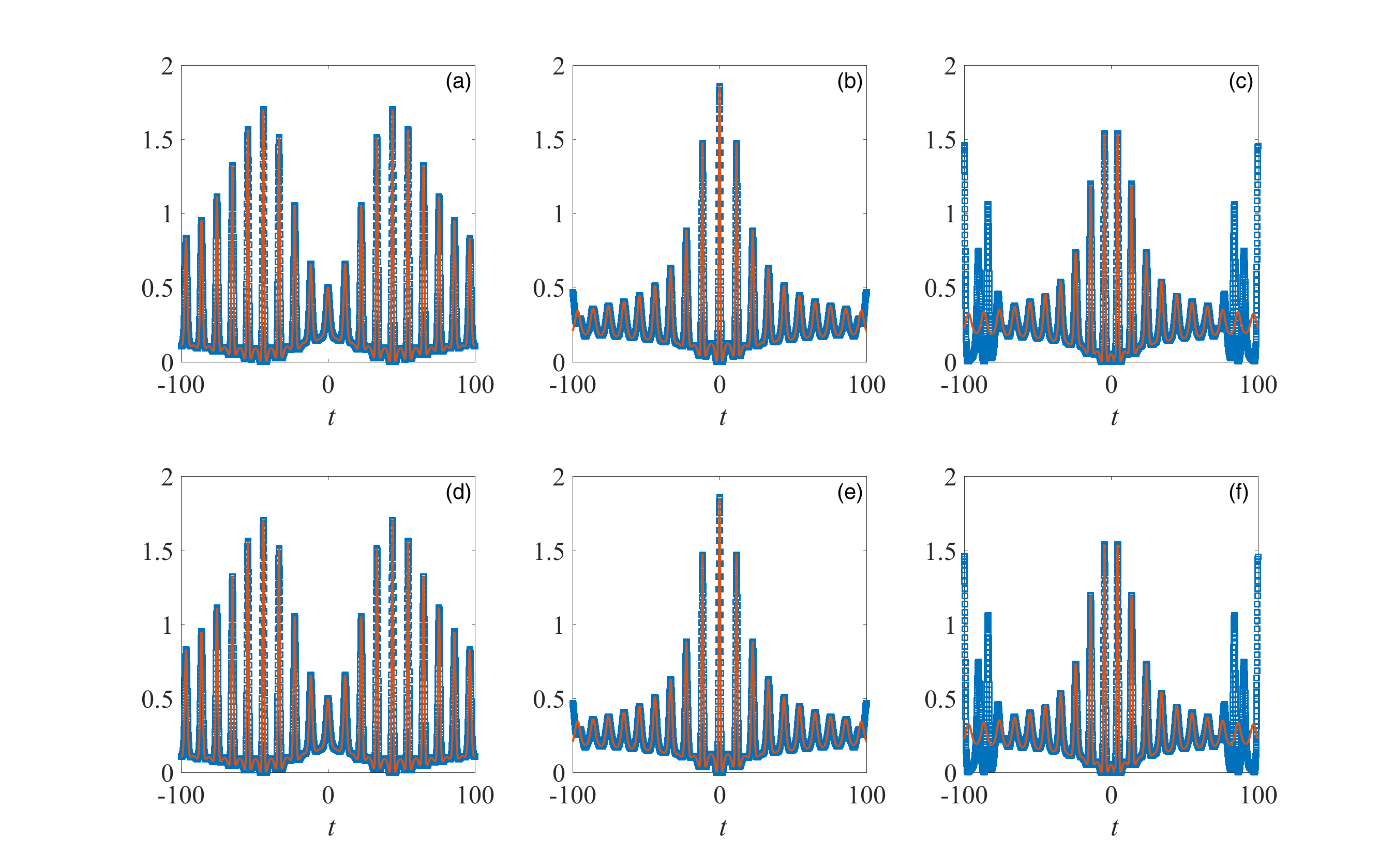}\\
   \caption{(Color Online)  Numerical simulation  of the triangular  pattern in central region of an order-2  {\it b-positon} $|q^{[2]}_{\rm{{\it b\mbox{-}positon}}}|^2$ with $s_0=0,s_1=50,a=0,c=\frac{1}{2},\lambda_1=\frac{2}{5}i$. Red line  denotes the theoretical (and exact) result, and square point denotes the result of numerical simulation.  The significant discrepancies occur
    at the two ends of period due to the reflection of simulation.
  (a) The initial signal of triangular pattern without noise at $z=-7.43$,
  (b) The observe signal of triangular pattern without noise at $z=-4.81$,
  (c) The observe signal of triangular pattern without noise at $z=4.42$,
  (d) The initial signal of triangular pattern with noise at $z=-7.43$ (SNR=100),
  (e) The observe signal of triangular pattern with noise at $z=-4.81$ (SNR=100),
  (f) The observe signal of triangular pattern with noise at $z=4.42$ (SNR=100).
}
\label{nscurvtriangular}
\end{figure}

\clearpage

\[\]


\begin{figure}[H]
  \centering
  \includegraphics[scale=.4]{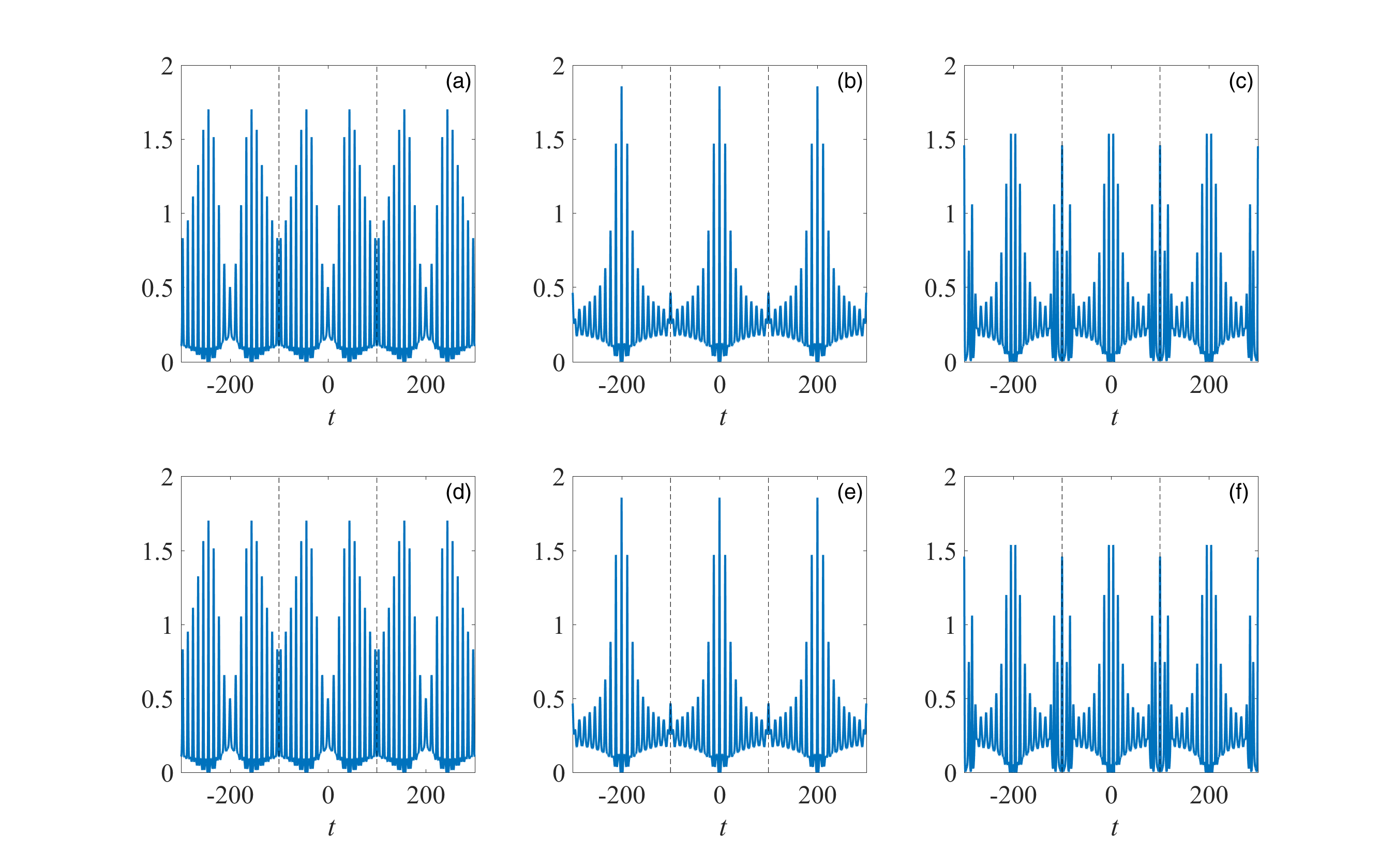}\\
   \caption{(Color Online)
    Numerical simulation of three periods from periodic extension of figure \ref{nscurvtriangular}. In order to get recognizable curve, theoretical results are not added.
  (a) The initial signal of triangular pattern without noise at $z=-7.43$,
  (b) The observe signal of triangular pattern without noise at $z=-4.81$,
  (c) The observe signal of triangular pattern without noise at $z=4.42$,
  (d) The initial signal of triangular pattern with noise at $z=-7.43$ (SNR=100),
  (e) The observe signal of triangular pattern with noise at $z=-4.81$ (SNR=100),
  (f) The observe signal of triangular pattern with noise at $z=4.42$ (SNR=100).
}
\label{nscurvtriangularextension}
\end{figure}

\clearpage
\subsection{The demonstration of instability for the order-2 {\it b-positon} by numerical simulation}
In this appendix, we use  numerical figures to show the increasing trend of the instability for the order-2 {\it b-positon} when $\lambda_1\rightarrow \lambda_0$.
\begin{figure}[h]
  \centering
  \subfigure[]{\includegraphics[scale=.3]{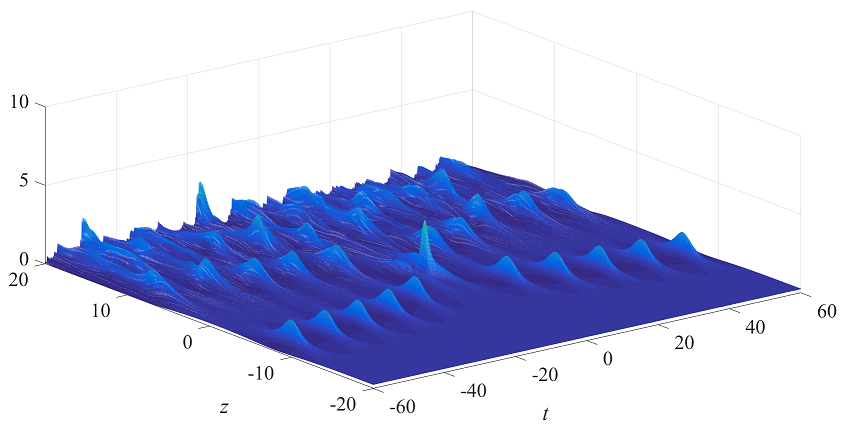}  \label{figinstability:a0}}
  \subfigure[]{\includegraphics[scale=.3]{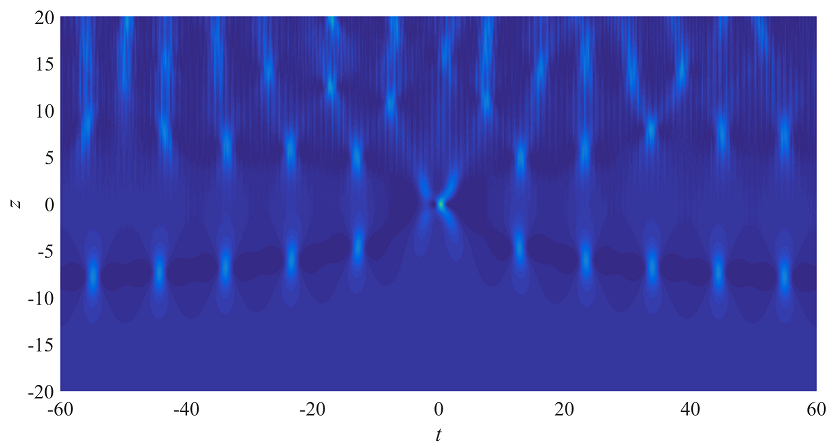}  \label{figinstability:b0}} \\
  \subfigure[]{\includegraphics[scale=.3]{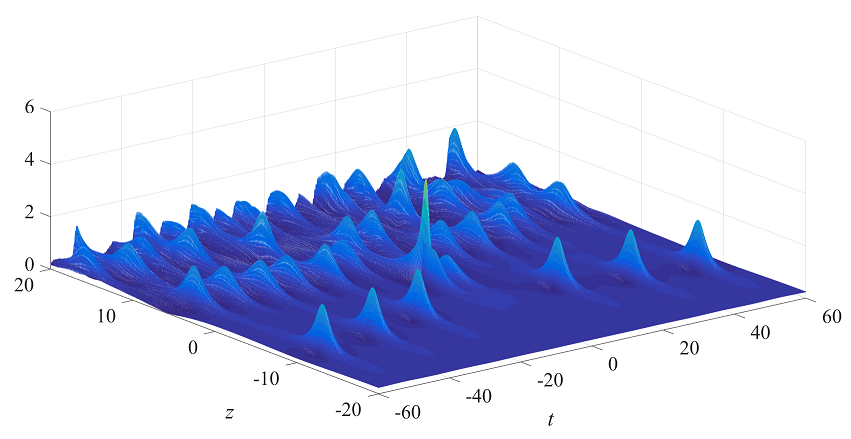}  \label{figinstability:c0}}
  \subfigure[]{\includegraphics[scale=.3]{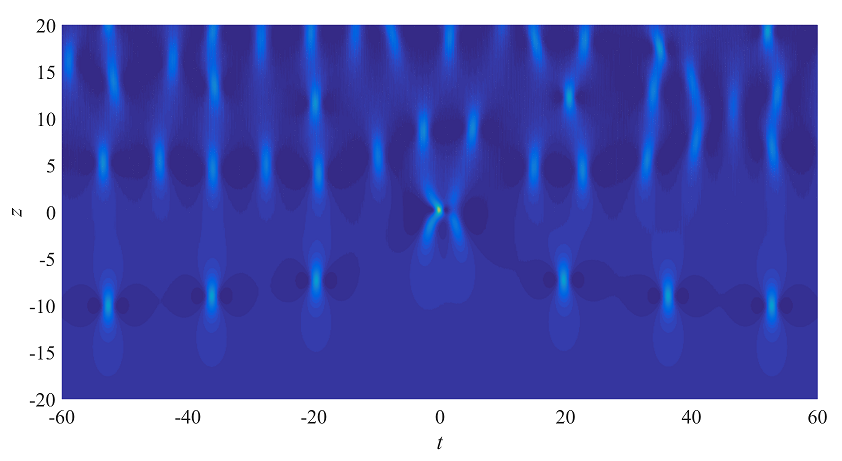}  \label{figinstability:d0}} \\
  \subfigure[]{\includegraphics[scale=.3]{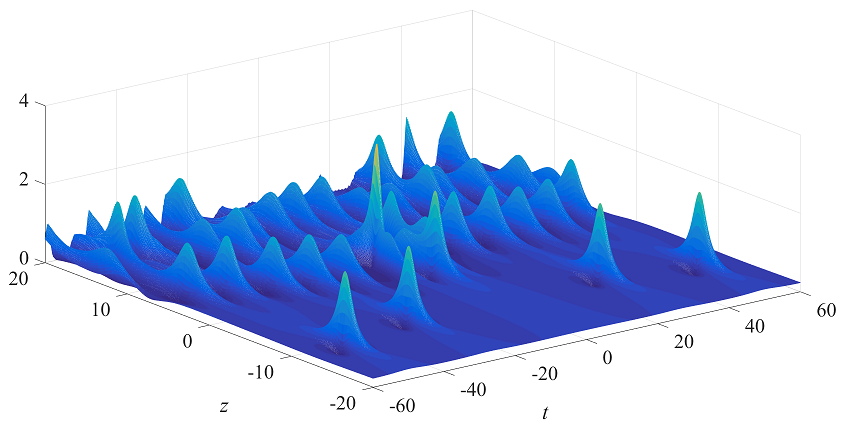}  \label{figinstability:e0}}
  \subfigure[]{\includegraphics[scale=.3]{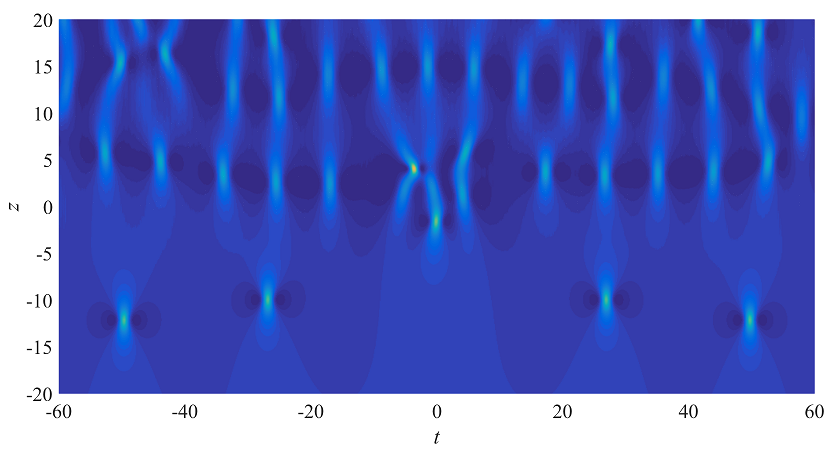}  \label{figinstability:f0}}\\
  \subfigure[]{\includegraphics[scale=.3]{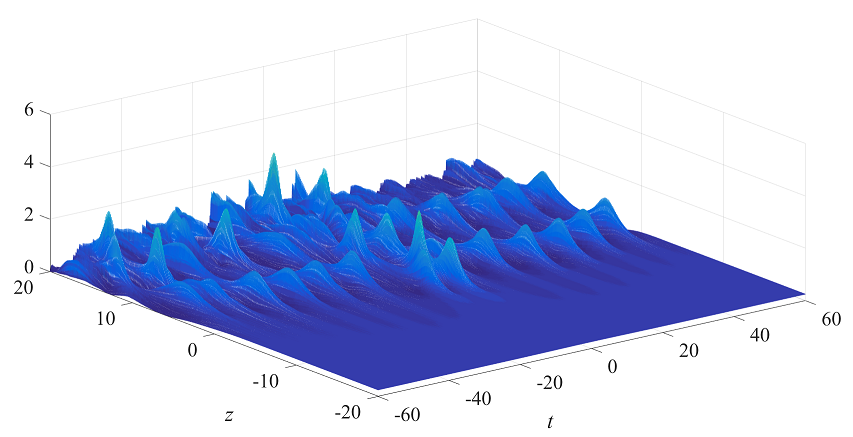}  \label{figinstability:g0}}
  \subfigure[]{\includegraphics[scale=.38]{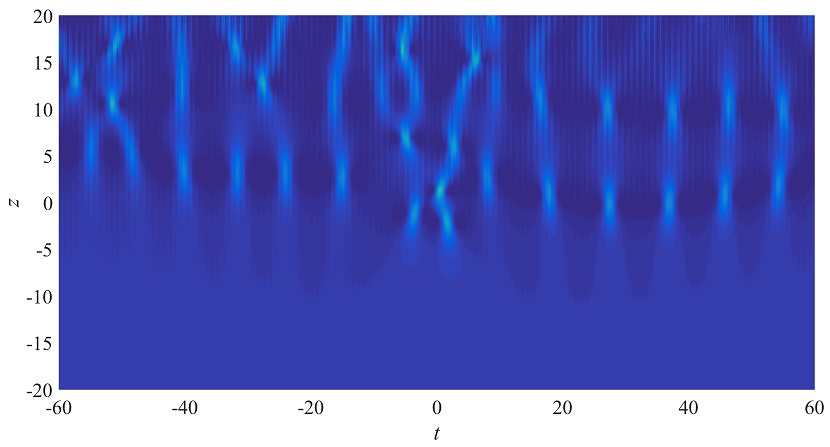}  \label{figinstability:h0}}
  \caption{(Color Online)  The  increasing  trend of instability for an order-2 {\it b-positon} $|q_{\rm{{\it b\mbox{-}positon}}}^{[2]}|^2$ with SNR=100 and parameters  $\lambda_0=-\frac{a}{2} + ic, a=0, c=\frac{1}{2},  s_0=0,s_1=0$.  From top to bottom, this {\it b-positon} is approaching an order-2 rogue wave as $\lambda_1$ tends to $\lambda_0$, which   corresponds to  the left column of  figure \ref{fig2}. The right column is the corresponding density plot of the left.
   The  parameter  $\lambda_1$ of {\it b-positons} is  $\frac{2}{5}i$ in (a,b), $\frac{6}{13}i$ in (c,d), $\frac{12}{25}i$ in (e,f), and $\frac{1}{2}i$ in (g,h).
  }
  \label{figinstability}
\end{figure}

\end{document}